\documentclass{aa}
\usepackage{natbib}
\usepackage{graphicx,psfig}
\usepackage{amssymb}
\begin{document}

\title{`Thermal' SiO radio line emission towards M-type AGB stars:\\ a probe
of circumstellar dust formation and dynamics\thanks{Based on observations
using the SEST at La Silla, Chile, the 20\,m
telescope at Onsala Space Observatory,
Sweden, the JCMT on Hawaii, and the IRAM 30\,m telescope at Pico
Veleta, Spain.}}

\titlerunning{SiO thermal emission from M-type AGB stars}

\author{D.~Gonz{\'a}lez Delgado\inst{1} \and H.~Olofsson\inst{1}
\and F.~Kerschbaum\inst{2} \and F.~L.~Sch\"oier\inst{1,3} \and M.~Lindqvist\inst{4} 
\and M.~A.~T.~Groenewegen\inst{5}}

\institute{Stockholm Observatory, AlbaNova, SE-10691 Stockholm, Sweden
\and Institut f\"ur Astronomie, T\"urkenschanzstrasse 17, 1180 Wien, Austria
\and Leiden Observatory, PO Box 9513, NL-2300 RA Leiden, The Netherlands
\and Onsala Space Observatory, SE--43992 Onsala, Sweden
\and Instituut voor Sterrenkunde, PACS-ICC, Celestijnenlaan 200B, 3001
Leuven, Belgium}

\offprints{H.~Olofsson (hans@astro.su.se)}
\date{Received 28 January 2003 / Accepted 3 July 2003}

\abstract{An extensive radiative transfer analysis of circumstellar
SiO `thermal' radio line emission from a large sample of M-type AGB
stars has been performed.  The sample contains 18 irregulars of type
Lb (IRV), 7 and 34 semiregulars of type SRa and SRb (SRV),
respectively, and 12 Miras.  New observational data, which
contain spectra of several ground vibrational state SiO rotational lines,
are presented.  The detection rate
was about 60\% (44\% for the IRVs, and 68\% for the SRVs). SiO
fractional abundances have been determined through radiative transfer
modelling.  The abundance distribution of the IRV/SRV sample has a
median value of 6$\times$10$^{-6}$, and a minimum of
2$\times$10$^{-6}$ and a maximum of 5$\times$10$^{-5}$.  The high mass-loss 
rate Miras have a much lower median abundance, $\lesssim$10$^{-6}$.  
The derived SiO abundances are in all
cases well below the abundance expected from stellar atmosphere
equilibrium chemistry, on average by a factor of ten.  In addition, there is a
trend of decreasing SiO abundance with increasing mass-loss rate.  This
is interpreted in terms of depletion of SiO molecules by the formation
of silicate grains in the circumstellar envelopes, with an efficiency
which is high already at low mass-loss rates and which increases with 
the mass-loss rate. The high mass-loss rate Miras
appear to have a bimodal SiO abundance distribution, a low
abundance group (on average 4$\times$10$^{-7}$) and a high abundance
group (on average 5$\times$10$^{-6}$).  The estimated SiO envelope
sizes agree well with the estimated SiO photodissociation radii using an
unshielded photodissociation rate of 2.5\,$\times$\,10$^{-10}$\,s$^{-1}$.
The SiO and CO radio line
profiles differ in shape.  In general, the SiO
line profiles are narrower than the CO line profiles, but they have
low-intensity wings which cover the full velocity range of the CO
line profile.  This is interpreted as partly an effect of selfabsorption
in the SiO lines, and partly (as has been done also by others) as due
to the influence of gas acceleration in the region which produces a
significant fraction of the SiO line emission.  Finally, a number of
sources which have peculiar CO line profiles are discussed from the
point of view of their SiO line properties. \keywords{Stars: AGB and post-AGB --
Circumstellar matter -- Stars: mass-loss -- Stars: late-type -- Radio lines: stars}}

\maketitle

\section{Introduction}

The atmospheres of and the circumstellar envelopes (CSEs) around
Asymptotic Giant Branch (AGB) stars are regions where many different
molecular species and dust grains form efficiently.  The molecular and
grain type setups are to a large extent determined by the C/O-ratio of
the central star.  For instance, SiO is formed in the extended
atmospheres of both M-type [C/O$<$1; O-rich] and C-type [C/O$>$1] AGB stars,
but its abundance is much higher in the former.  Therefore, the SiO
`thermal' line emission (i.e., rotational lines in the $v$=0 state;
the term `thermal' is used here to distinguish the $v$=0 state
emission from the strong maser line emission from vibrationally
excited states) is particularly strong towards M-stars, with the
intensity of e.g. the $J$=2$\rightarrow$1 line comparable to, or even
stronger than, that of the CO $J$=1$\rightarrow$0 emission.
Nevertheless, the initial observations of SiO thermal radio line
emission from AGB-CSEs \citep{lambvand78, wolfcarl82} and
their interpretation \citep{morretal79} suggested
circumstellar SiO abundances (several) orders of magnitude lower than
those expected from the chemical equilibrium models \citep{tsuj73}.

Over the years the observational basis has improved considerably
\citep{bujaetal86, bujaetal89, bieglatt94, biegetal98,
biegetal00, olofetal98a}, and even some
interferometer data exist \citep{lucaetal92, sahabieg93}.  
These data suggest that the SiO line
emission originates in two regions, one close to the star with a high
SiO abundance, and one extended region with a low SiO abundance.
The relative contributions to
the SiO line emission from these two regions depend on the mass-loss
rate.

This structure has been interpreted as due to accretion of SiO onto
dust grains \citep{bujaetal89, sahabieg93}.  After the grains nucleate 
near the stars, they grow in
part because of adsorption of gas-phase species.  In O-rich CSEs,
refractory elements like Si, together with O, are very likely the main
constituents of the grains, which are identified through the 9 and
18~$\mu$m silicate features in the infrared spectra of the stars
\citep{forretal75, pegopapo85}.  Therefore, molecules like SiO are expected to be
easily incorporated into the dust grains.  As a result, the SiO gas
phase abundance should fall off with increasing distance from the star
as SiO molecules in the outflowing stellar wind are incorporated into
the grains.  The depletion process is, however, quite uncertain since
it does not proceed at thermal equilibrium.  Eventually,
photodissociation destroys all of the remaining SiO molecules.

The grain formation is important not only for the chemical composition
of the CSE, but also because it affects its dynamical state [the
radiation pressure acts on the grains which are dynamically coupled to
the gas, e.g., \citet{kwok75}].  The SiO radio line profiles are
narrower than those of CO and have mostly Gaussian-like shapes (e.g.,
\citet{bujaetal86, bujaetal89}), a fact
suggesting that the SiO line emission stems from the inner regions of
the CSEs, where grain formation is not yet complete and where the
stellar wind has not reached its terminal expansion velocity.  This result
is corroborated by interferometric observations which show that the
size of the SiO line emitting region is independent of the line-of-sight
velocity \citep{lucaetal92}.  Lucas et al.  explained this
as a result of a rather extended acceleration region.  However, \citet{sahabieg93}, 
using a more detailed modelling, were able
to explain both the line profiles and the brightness distributions
with a `normal' CSE, i.e., with a rather high initial acceleration.

Therefore, `thermal' SiO radio line emission is a useful probe of the
formation and evolution of dust grains in CSEs, a complex phenomenon
that is yet not fully understood, as well as the CSE dynamics.

In this paper we
present a detailed study of SiO radio line emission from the CSEs
of a sample of M-type AGB stars.  The sample includes irregular
(IRVs), semiregular (SRVs) and Mira (M) variables.  The IRVs and SRVs
have already been studied in circumstellar CO radio line emission
\citep{olofetal02}, yielding estimates of the
stellar mass-loss rates.  Using these estimates a radiative transfer
modelling of the SiO radio line emission is performed.  A complete
analysis of the circumstellar CO and SiO line emission is done for the
Mira sub-sample.

\section{Observations of the IRV/SRV sample}

\subsection{The IRV/SRV sample}

The sample contains all the M-type IRVs and SRVs detected
in circumstellar CO radio line emission by \citet{kersolof99} 
and \citet{olofetal02}.  The
original source selection criteria are described in \citet{kersolof99}, 
but basically these stars are the
brightest 60~$\mu$m-sources (IRAS $S_{60}$ typically above 3\,Jy, with
IRAS quality flag 3 in the 12, 25, and 60~$\mu$m bands) that appear as
IRVs or SRVs in the General Catalogue of Variable Stars [GCVS4;
\citet{kholetal90}].  The detection rate of circumstellar CO
was rather high, about 60\% (\citet{olofetal02}; 69
stars detected).  The basic properties of the stars are listed in
\citet{kersolof99} and \cite{olofetal02}.

The distances, presented in Table~\ref{t:SiOmodelresults}, were
derived using an assumed bolometric luminosity of 4000 L$_\odot$ for
all stars.  We are aware of the fact that such a distance estimate have a
rather large uncertainty for an individual object but it is adequate for a
statistical study of a sample of stars (see discussion by \citet{olofetal02}).  
The apparent bolometric fluxes were
obtained by integrating the spectral energy distributions ranging from
the visual data over the near-infrared to the IRAS-range \citep{kershron96}.

\subsection{The observing runs}

The SiO ($v$=0, $J$=2$\rightarrow$1; hereafter all SiO transitions are in the ground
vibrational state) data were obtained using the 20\,m telescope at
Onsala Space Observatory (OSO) and the 15\,m Swedish-ESO Submillimetre
Telescope (SEST) on La Silla, Chile.  At SEST, a sizable fraction of
the stars were observed also in the SiO $J$=3$\rightarrow$2 line, and four
additional sources were observed with the IRAM 30\,m telescope at Pico Veleta,
Spain, in this line.  The
higher-frequency lines, $J$=5$\rightarrow$4 and 6$\rightarrow$5,
were observed towards 10 and 3 stars, respectively. The observing
runs at OSO were made over the years 1993 to 2000, at SEST over
the years 1992 to 2003, and at IRAM between October 18 and 22 in 1997.  
Telescope and receiver data are given in
Table~\ref{t:telescopes}.  $T_{\rm rec}$ and $\eta_{\rm mb}$ stand for
the representative noise temperature of the receiver (SSB) and the
main beam efficiency of the telescope, respectively.

Two filterbanks at OSO (256$\times$250\,kHz, and 512$\times$1\,MHz),
two acousto-optical spectrometers at SEST (86\,MHz bandwidth with
43\,kHz channel separation, and 1\,GHz bandwidth with 0.7\,MHz channel
separation), and a 1\,MHz filter bank at IRAM were used as spectrometers.  
Dual beam switching (beam
throws of about 11\arcmin), in which the source was placed alternately
in the two beams, was used to eliminate baseline ripples at OSO and SEST, while
a wobbler switching with a throw of 150$\arcsec$ in azimuth was used at
IRAM. Pointing and focussing were checked every few hours. The line
intensities are given in the main beam brightness temperature scale
($T_{\rm mb}$), i.e., the antenna temperature has been corrected for
the atmospheric attenuation (using the chopper wheel method) and
divided by the main beam efficiency.

\begin{table}
     \caption[]{Data on telescopes and receivers}
      \label{t:telescopes}
      \[
      \begin{tabular}{ccccc}
      \hline
      Telescope & Frequency & Beamwidth & $T_{\rm rec}$ & $\eta_{\rm mb}$ \\
                & [MHz]     & [$\arcsec$] & [K] \\
      \hline
      OSO   &  86847 & 42 & 150  & 0.55 \\
      SEST  &  86847 & 57 & 100  & 0.75 \\
      SEST  & 130269 & 39 & 120  & 0.65 \\
      IRAM  & 130269 & 18 & 150  & 0.58 \\
      SEST  & 217105 & 25 & 600  & 0.55 \\
      SEST  & 260518 & 21 & 800  & 0.45\\
      \hline
      \end{tabular}
      \]
\end{table}

\subsection{Observational results}

A total of 60 stars were observed in circumstellar SiO line emission
(i.e., about 85\% of the stars detected in circumstellar CO): 34 stars
were detected in the SiO $J$=2$\rightarrow$1 line, 21 in the
$J$=3$\rightarrow$2 line, and 3 in the $J$=5$\rightarrow$4 and
$J$=6$\rightarrow$5 lines.
Clear detections of SiO lines were obtained towards 36 sources, i.e.,
the detection rate was about 60\%: 8 IRVs (detection rate 44\%) and 28
SRVs (detection rate 68\%) were detected.
Tables~\ref{t:srvirvobsresultsI} and \ref{t:srvirvobsresultsII} in the
Appendix list all our SiO observations. The names in the GCVS4 and
the IRAS-PSC are
given.  The first letter of the code denotes the observatory
({\bf I}RAM, {\bf O}SO, or {\bf S}EST), the rest the transition observed. 
Another code reflects the `success' of the observation ({\bf
D}etection, {\bf N}on-detection).

The stellar velocity is given with respect to the heliocentric
($v_{\rm hel}$) and LSR frame [$v_{\rm LSR}$; the Local Standard of
Rest is defined using the standard solar motion (B1950.0):
$v_\odot$=\,20\,km\,s$^{-1}$, $\alpha_\odot$=\,270.5$^{\circ}$,
$\delta_\odot$=\,+30$^{\circ}$].  The stellar velocity, the expansion
velocity, and the main beam brightness temperature were obtained by
fitting the function $T_{\rm mb} [1-((v_{*}-v_{\rm z})/v_{\rm
e})^2]^{\gamma}$ to the line profile.  The integrated intensity,
$I$=\,$\int T_{\rm mb}$\,$dv$, is obtained by integrating the line
intensities over the line profile.  The uncertainty in $I$ varies with
the S/N-ratio, but we estimate that it is on avarage $<$15\%.  To this
should be added an estimated uncertainty in the absolute calibration
of about 20\%.  For a non-detection an upper limit to $I$ is estimated
by measuring the peak-to-peak noise ($T_{\rm pp}$) of the spectrum
with a velocity resolution reduced to 15\,km\,s$^{-1}$ and calculating
$I$\,=\,15$T_{\rm pp}$.  The Q-column gives a quality ranking: 5 (not
detected), 4 (detection with very low S/N-ratio $\lesssim$3), 3
(detection, low S/N-ratio $\approx$5), 2 (detection, good S/N-ratio
$\approx$ 10), and 1 (detection, very good S/N-ratio $\gtrsim$15). 
Finally, in cases of complex velocity profiles the measured component
is indicated in the form {\bf b}=broad, {\bf n}=narrow, {\bf b}+{\bf
n}=total.

All the spectra are shown in Figs.~\ref{f:siolb} to \ref{f:siosr3}. The
velocity scale is given in the heliocentric system.  The velocity
resolution is reduced to 0.5\,km\,s$^{-1}$, except for some low
S/N-ratio spectra where a resolution of 1\,km\,s$^{-1}$, or even
2\,km\,s$^{-1}$, is used, and for some low expansion velocity sources
for which 0.25\,km\,s$^{-1}$ is used.

\section{The Mira sample}

In order to make a more extensive study of circumstellar SiO line
emission in the CSEs of M-type AGB-stars, a sample of 12 Mira
variables with higher mass-loss rates was added.  The
distances are obtained using the period--luminosity relation of
\citet{whitetal94}.  Through modelling of their
circumstellar CO radio line emission (Sect.~\ref{s:mirascomodel}), we
determined that 4 of the Miras have very high mass-loss rates
($\gtrsim$\,10$^{-5}$\,M$_\odot$\,yr$^{-1}$), 6 are intermediate to high mass
loss rate objects ($\gtrsim$\,10$^{-6}$\,M$_\odot$\,yr$^{-1}$) and 2
are low mass-loss rate sources
(a few 10$^{-7}$\,M$_\odot$\,yr$^{-1}$).

For this sample data has been gathered from a number of sources. 
The CO($J$=1$\rightarrow$0) data were taken from \citet{olofetal98a}, while the CO($J$=2$\rightarrow$1, $J$=3$\rightarrow$2, and
$J$=4$\rightarrow$3) data were obtained from the archive of the James
Clerk Maxwell Telescope on Mauna Kea, Hawaii.  The JCMT data
are taken at face value after converting to the main beam brightness
scale.  However, in the cases where there are more
than one observation available, the derived line intensities are
generally consistent within $\pm$20\% (as was found also by \citet{schoolof01}).  The
SiO($J$=2$\rightarrow$1) data were obtained from \citet{olofetal98a}.  
The SiO $J$=5$\rightarrow$4 line was observed in four objects, and
the $J$=6$\rightarrow$5 line in one object using SEST with
the same observational equipment and procedure as described above.

The relevant observational results are
summarized in Table~\ref{t:mirasobsresults}, and the SiO spectra are shown in
Fig.~\ref{f:siomira}.  The names in the
GCVS4 and the IRAS-PSC are given.  The first letter of the code
denotes the observatory ({\bf J}CMT, {\bf O}SO, or {\bf S}EST), the
rest the transition observed.

\section{Modelling of circumstellar line emission}

Apart from presenting new observational results on thermal SiO radio
line emission from AGB-CSEs a rather detailed modelling of the
emission will be performed.  In some senses this is a more difficult
enterprise than the CO line modelling.  The SiO line emission predominantly
comes from a region closer to the star than does the CO line emission,
and this is a region where the observational constraints are poor.  The SiO
excitation is also normally far from thermal equilibrium with the gas
kinetic temperature, and radiative excitation plays a larger role
(hence the term `thermal' is really not appropriate).  Finally, there exists
no detailed chemical model for calculating the radial SiO abundance
distribution.  These effects make the SiO line modelling much more uncertain,
and dependent on a number of assumptions.

\begin{figure}
     \centerline{\psfig{file=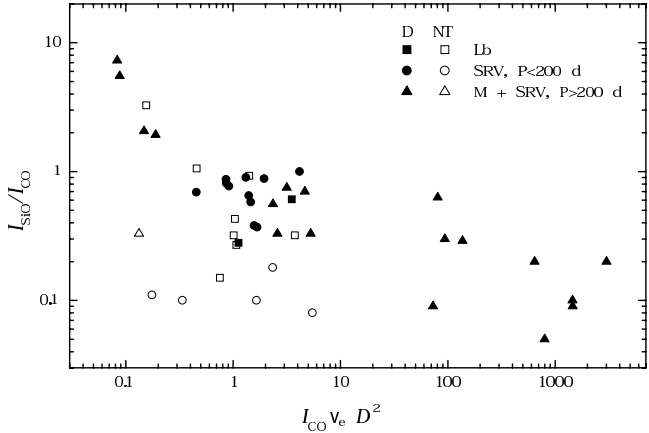,width=8.5cm}}    
     \caption{The line intensity ratio
     $I$(SiO,$J$=2$\rightarrow$1)/$I$(CO,$J$=1$\rightarrow$0) as a function
     of a mass-loss rate measure [detections are shown as filled symbols,
     while tentative and negative results are shown as open symbols; see
     \citet{olofetal98a}) for more details]}
     \label{f:sioco}
\end{figure}

The aim is to investigate to what extent the thermal SiO line emission
is a useful probe of e.g. the dust formation and the CSE dynamics.
There are observational indications that this is the case but the
interpretation is normally not straightforward. As an example,
\citet{olofetal98a} found that the line intensity ratio
$I$(SiO,$J$=2$\rightarrow$1)/$I$(CO,$J$=1$\rightarrow$0) decreases
markedly as a function of a mass-loss rate measure.  Their results are
reproduced here, but now including all stars of our IRV/SRV and Mira
samples, Fig.~\ref{f:sioco}.  A straightforward interpretation would
be that the SiO abundance decreases with mass-loss rate due to
increased depletion efficiency and hence this limits severely the SiO
line strength.  However, excitation may play an important role here,
both for SiO and CO, and a detailed modelling is required.

\subsection{The method}

In order to model the circumstellar SiO line emission a non-LTE
radiative transfer code based on the Monte Carlo method has been used
\citep{bern79}. It has
been previously used to model circumstellar CO radio line emission in
samples of both C- \citep{schoolof00, schoolof01, schoetal02} and O-rich 
\citep{olofetal02} AGB-CSEs, and also to model the HCN and CN line emission
from a limited number of C-rich AGB-CSEs \citep{lindetal00}.

\subsection{The SiO molecule}

In the excitation analysis of SiO 50 rotational levels in
both the ground and the first excited vibrational state are considered.
The energy levels of this linear rotor are calculated using the
molecular constants from \citet{molletal91}.
The radiative rates are calculated using the dipole moment from
\citet{raymetal70}.  Collisional deexcitation rates have
been calculated by \citet{turnetal92}) in the temperature
range 20--300\,K and up to $J$$=$20.  The original data set has been
extrapolated in temperature and to include levels up to $J$$=$50
(Sch\"oier et al., in prep.).

\subsection{The circumstellar model}

The CSEs around AGB-stars are intricate systems where an interplay
between different chemical and physical processes takes place. This
makes the modelling of circumstellar radio line emission a quite
elaborate task. In the analysis presented here, a
relatively simple, yet realistic, model for the geometry and
kinematics of the CSEs has been adopted.  Below follows a short
description of the main features of the circumstellar model.  For more
details we refer to \citet{schoolof01} and
\citet{olofetal02}.

A spherically symmetric geometry of the CSE is adopted.  The mass loss
is assumed to be isotropic and constant with time.  The gas expansion
velocity is assumed to be constant with radius.  There is a
possibility that neither the mass-loss rate nor the expansion velocity
are constant in the regions of interest here.  This should be kept in
mind when interpreting the results.  There is growing evidence for
mass-loss modulations of AGB-stars on a time scale of about 1000 yr
\citep{maurhugg00, mareetal01a,
fongetal03}, and the CO line emission comes from a much
larger region than that of the SiO lines, and hence averages over a
longer time span.  Furthermore, the SiO line
emission comes from the inner part of the CSE, where it is likely that
the gas has not fully reached the terminal velocity.  We have not
allowed for the presence of gas acceleration nor a time-variable mass
loss in the modelling in order to limit the number of free parameters.

The inner boundary of the CSE was set to 1\,$\times$\,10$^{14}$\,cm
($\approx$3R$_*$).  This parameter is specially important in
the case of SiO where radiative excitation is expected to play a role.  A
turbulent velocity of 0.5\,km\,s$^{-1}$ is assumed throughout the
entire CSE (see discussion by \citet{olofetal02}).
The outer boundaries of the molecular abundance distributions are, for
both CO and SiO, determined by photodissociation due to the
interstellar UV radiation field.  For CO we use the modelling of 
\citet{mamoetal88}.  The procedure for SiO is presented in
Sect.~\ref{s:siosize}.

The radiation field is provided by two sources. The central radiation
emanates from the star.  This radiation was estimated from a fit to
the spectral energy distribution (SED) by assuming two blackbodies,
one representing the direct stellar radiation and one the
dust-processed radiation \citep{kershron96}. 
In the case of optically thin dust CSEs the stellar blackbody temperature derived in
this manner is generally about 500\,K lower than the effective
temperature of the star.  The dust mass-loss rates of the IRV/SRVs are low
enough that the dust blackbody can be ignored.  For the sample of Mira
variables both blackbodies were used, since for these high mass-loss
rate stars, the excitation of the SiO molecules may be affected by
dust emission.  The second radiation field is provided by the cosmic
microwave bakground radiation at 2.7\,K.

In the SiO line modelling the gas kinetic temperature
law derived in the modelling of the circumstellar CO radio line
emission was used.  This is reasonable since the SiO line emission
contributes very little to the cooling of the gas. However, the SiO line 
emission comes mainly from the inner CSE, where the CO lines do
not put strong constraints on the temperature, and where other coolants, 
specifically H$_2$O, may be important. We estimate though that the
kinetic tempartures used in our modelling are not seriously wrong.
In addition,  for at least the
lower mass-loss rates the SiO molecule is mainly radiatively excited,
and hence the exact gas kinetic temperature law and the collisional
rate coefficients play only a minor role, see Sect.~\ref{s:dependence}.

In Sect.~\ref{s:dependence} some implications of these assumptions are
discussed.

The mass-loss rates for the sample of IRV/SRVs were already presented
in \citet{olofetal02}.  They were derived through
modelling of circumstellar CO radio line observations.  A median mass
loss rate of 2$\times$10$^{-7}$\,M$_\odot$\,yr$^{-1}$ was found for this
sample.  These mass-loss rate estimates are expected to be accurate to
within a factor of a few for an individual object.  Nevertheless, they
are probably the best mass-loss rate estimates for these types of
objects [also in agreement with the mass-loss rate estimates by
\citet{knapetal98} for five sources in common].

The modelling of the circumstellar CO radio line emission for the Mira
sample is presented in this paper.  The same approach as in
\citet{olofetal02} has been used, i.e., the energy balance
equation is solved simultaneously with the CO excitation.  A
number of (uncertain) parameters describing the dust are introduced.
They are grouped in a global parameter, the $h$-parameter, which is given
by
\begin{equation}
h = \left[\frac{\psi}{0.01}\right]
\left[\frac{2.0\,\mathrm{g\,cm}^{-3}}{\rho_\mathrm{gr}}\right]
\left[\frac{0.05\,\mu\mbox{m}}{a_\mathrm{gr}}\right],
\label{h}
\end{equation}
where $\psi$ is the dust-to-gas mass ratio, $\rho_\mathrm{gr}$ the dust
grain density, and $a_\mathrm{gr}$ its radius.  This parameter is
particularly important for the heating due to gas-grain collisions. The
normalized values are the ones used to fit the CO radio line emission
of \object{IRC+10216} using this model
\citet{schoolof01}, i.e., $h$=1 for this object.
\citet{schoolof01} found that on average $h$=0.2 for the lower
luminosity sources (below 6000\,L$_{\odot}$; and $h$=0.5 for the more
luminous sources) in their sample of bright carbon stars and 
\citet{olofetal02} found $h$=0.2 for their sample of M-type IRV/SRVs.
In addition, following \citet{olofetal02}
we use (in the gas-grain drift heating term) a flux-averaged momentum transfer
efficiency from dust to gas, $Q_{\rm p,F}$, equal to 0.03 independent
of the mass-loss rate, and adopt a CO abundance with respect to H$_2$
of 2\,$\times$\,10$^{-4}$.  The latter may very well be an
underestimate for these high mass-loss rate objects (see below).

\subsection{Dependence on parameters}
\label{s:dependence}

\begin{table*}
\caption[ ]{The effect on the integrated model SiO intensities (in percent),
due to changes in various parameters.  Three model stars with mass
loss rate and gas expansion velocity characteristics typical for our
samples are used.  They lie at a distance of 250\,pc, and have
luminosities of 4000\,L$_{\odot}$ (the model stars with mass-loss
rates of 10$^{-7}$ and 10$^{-6}$\,M$_{\odot}$\,yr$^{-1}$) and 8000\,L$_{\odot}$ (the
model star with a mass-loss rate of 10$^{-5}$\,M$_{\odot}$\,yr$^{-1}$), and blackbody
temperatures of 2500\,K.  The nominal CSE parameters are $h$=0.2 (for
the lowest mass-loss rate model star) and $h$=0.5 (for the other two
model stars), $r_{\rm i}$=2$\times$10$^{14}$\,cm, $v_{\rm
t}$=0.5\,km\,s$^{-1}$, and $f_{\rm SiO}$=5$\times$10$^{-6}$. The size of the 
SiO envelope, $r_{\rm e}$, is given by Eq.(\ref{e:re}) for the given mass-loss rate. The
SiO $J$=2$\rightarrow$1, $J$=3$\rightarrow$2,
$J$=5$\rightarrow$4, and $J$=6$\rightarrow$5 lines are
observed with beam widths of 57$\arcsec$, 38$\arcsec$,
23$\arcsec$, and 19$\arcsec$, respectively (appropriate for a 15\,m telescope).
The model integrated line intensities, $I$ in K\,km\,s$^{-1}$, are given for the nominal
parameters. For comparison, also the integrated CO line intensities for the model stars are
given [for a 15\,m telescope; 45$\arcsec$ ($J$=1$\rightarrow$0;
$I$\,=\,0.14, 3.8, and 45\,K\,km\,s$^{-1}$ for 10$^{-7}$,10$^{-6}$, and
10$^{-5}$\,M$_{\odot}$\,yr$^{-1}$, respectively), 23$\arcsec$
($J$=2$\rightarrow$1), 15$\arcsec$ ($J$=3$\rightarrow$2), 9$\arcsec$
($J$=5$\rightarrow$4), 7$\farcs$5 ($J$=6$\rightarrow$5)]}
\begin{flushleft}
\begin{tabular}{crrrrrrrrrrrrrrr}
\hline
\noalign{\smallskip}
   & &
\multicolumn{4}{c}{10$^{-7}$\,M$_{\odot}$\,yr$^{-1}$,
7\,km\,s$^{-1}$} & &
\multicolumn{4}{c}{10$^{-6}$\,M$_{\odot}$\,yr$^{-1}$,
10\,km\,s$^{-1}$} & &
\multicolumn{4}{c}{10$^{-5}$\,M$_{\odot}$\,yr$^{-1}$,
15\,km\,s$^{-1}$}
\\
\cline{3-6}
\cline{8-11}
\cline{13-16}
\multicolumn{1}{c}{Par.} &
\multicolumn{1}{c}{Change} &
\multicolumn{1}{r}{2$-$1} &
\multicolumn{1}{r}{3$-$2} &
\multicolumn{1}{r}{5$-$4} &
\multicolumn{1}{r}{6$-$5} &
&
\multicolumn{1}{r}{2$-$1} &
\multicolumn{1}{r}{3$-$2} &
\multicolumn{1}{r}{5$-$4} &
\multicolumn{1}{r}{6$-$5} &
&
\multicolumn{1}{r}{2$-$1} &
\multicolumn{1}{r}{3$-$2} &
\multicolumn{1}{r}{5$-$4} &
\multicolumn{1}{r}{6$-$5}
\\
\noalign{\smallskip}
\hline
\noalign{\smallskip}
$I_{\rm SiO}$ &  &0.11&0.30&0.71&0.90&   &1.3&2.7&5.5&6.8&  &10&20&39&48\\
\hline
\noalign{\smallskip}
$I_{\rm CO}$ &   &1.6&3.9&6.0&6.4&   &17&31&46&50&   &120&184&252&271\\
\hline
\noalign{\smallskip}
$f_{SiO}$       & $-$50\%    & $-$40&$-$37&$-$36&$-$36&   &$-$33&$-$30&$-$31&$-$32&   &$-$27&$-$24&$-$26&$-$27\\
                 & $+$100\%  & $+$70&$+$56&$+$44&$+$44&   &$+$46&$+$39&$+$40&$+$42&   &$+$29&$+$27&$+$36&$+$37\\
$L$             &  $-$50\%   & 0&$-$7&$-$14&$-$14&   &$-$8&$-$5&$-$4&$-$4&    &$-$10&$-$2&0&0\\
                &  $+$100\%  & 0&$+$11&$+$19&$+$21&   &$+$13&$+$14&$+$10&$+$9&   &$+$19&$+$8&$+$4&$+$3\\
$T_{\mathrm kin}$&  $-$33\%   & 0&0&$-$7&$-$8&  &$-$5&$-$9&$-$16&$-$19&  &$-$16&$-$21&$-$27&$-$31\\
                 &  $+$50\%   & 0&$+$4&$-$12&$+$7&  &$+$8&$+$11&$+$16&$+$18&  &$+$20&$+$24&$+$32&$+$35\\
$r_{\mathrm e}$ & $-$50\%    & $-$60&$-$56&$-$39&$-$33&   &$-$48&$-$38&$-$29&$-$25&   &$-$35&$-$33&$-$26&$-$22\\
                 & $+$100\%  & $+$110&$+$70&$+$32&$+$24&   &$+$53&$+$41&$+$22&$+$14&   &$+$41&$+$39&$+$18&$+$9\\
$r_{\mathrm i}$ & $-$50\%   & $+$10&$+$4&$+$2&$+$3&  &$+$2&$+$2&$+$1&0&   &$+$1&$-$1&$-$1&0\\
	        & $+$100\%  & 0&$-$4&$-$8&$-$11&   &$-$1&0&$-$3&$-$4&    &$+$1&$+$1&$-$1&$-$3\\
$v_{\mathrm t}$ & $-$50\%   & 0&0&$-$2&$-$1&   &$+$1&$+$1&$-$1&$+$2&    &$-$6&$-$6&$-$2&$-$2\\
	        & $+$100\%  & 0&$+$4&$+$2&$+$3&  &0&$+$3&$+$3&$+$2&    &$+$3&$+$3&$+$1&0\\
\noalign{\smallskip}
\hline
\noalign{\smallskip}
\end{tabular}
\end{flushleft}
\label{t:pardep}
\end{table*}

A sensitivity test has been performed in order to determine the
dependence of the calculated SiO line intensities on the assumed
parameters for a set of model stars.  They are chosen such that they
have nominal mass-loss rate and gas expansion velocity combinations
which are characteristic of our samples: a low mass-loss rate
(10$^{-7}$\,M$_{\odot}$\,yr$^{-1}$, 7\,km\,s$^{-1}$), an intermediate
mass-loss rate (10$^{-6}$\,M$_{\odot}$\,yr$^{-1}$, 10\,km\,s$^{-1}$),
and a high mass-loss rate (10$^{-5}$\,M$_{\odot}$\,yr$^{-1}$,
15\,km\,s$^{-1}$) model star.  They are placed at a distance of
250\,pc (a typical distance of the stars in the IRV/SRV sample).  We
have also taken nominal values for the luminosity ($L$=$4000\,$L$_\odot$
for the low and intermediate mass-loss rate model stars, and
$L$=$8000\,$L$_\odot$ for the high mass-loss rate model star), the
effective temperature ($T_{\rm bb}$=$2500$K), the $h$-parameter
($h$=0.2 for the low mass-loss rate model star, and $h$=0.5 for the
other two), the envelope inner radius
($r_i$=2\,$\times$\,10$^{14}$\,cm, which is twice the inner radius
used in the modelling), the turbulent velocity
($v_t$=0.5\,km\,s$^{-1}$), and the SiO abundance [$f_{\rm
SiO}$=5\,$\times$\,10$^{-6}$ (close to the median value for our
IRV/SRV sample, see below); throughout this paper the term abundance
means the fractional abundance with respect to H$_2$, the dominating
molecular species in the CSEs].  The SiO envelope outer radius is
calculated for each model star following the same relation that is
used in the modelling of the sample stars (see
Sect.~\ref{s:interferomdata}).  The SiO lines are observed with beam
widths characteristic of our observations.  All parameters (except the
mass-loss rate and expansion velocity) are changed by $-$50\% and
$+$100\% and the velocity-integrated line intensities are calculated. 
In order to check the effect of the $h$-parameter on the modelled
intensities the radial gas kinetic temperature law is scaled by
$-$33\% and $+$50\%.  The results are summarized in
Table~\ref{t:pardep} in terms of percentage changes.  To see how the
SiO/CO line intensity ratios vary with mass-loss rate, the CO line
intensities derived from the models with the nominal parameters are
also included.

Despite the fact that
the dependences are somewhat complicated there are some general
trends.  The line intensities are, in general, sensitive to changes in
the outer radius, but less so for the high-$J$ lines, a fact which is
more evident for the low mass-loss rate stars.  There is also a
dependence of all line intensities on the SiO abundance,
irrespective of the magnitude of the mass-loss rate.  These particular
dependences of the line intensities on the envelope outer radius and
the SiO abundance allowed us to derive envelope sizes for those stars
with multi-line observations (see Sect.~\ref{s:siomodelresults}).  The
line intensities are rather insensitive to a change in the kinetic
temperature.  Only the high-$J$ lines for high mass-loss rates show a
weak dependence on this parameter.  The dependence on the inner radius
is marginal, and so is the dependence on the turbulent velocity width
(as long as it is significantly smaller than the expansion velocity).

The dependece on luminosity is also weak, with only small
changes in high-$J$ line intensities for low mass-loss rates and in
low-$J$ line intensities for high mass-loss rates.  However, the
radiation field distribution may be of importance here, in particular
for the high mass-loss rate objects.  We have checked this for the high
mass-loss rate model star.  If half of the luminosity is put in a 750\,K
blackbody, the $J$=2$\rightarrow$1, $J$=3$\rightarrow$2,
$J$=5$\rightarrow$4, and $J$=6$\rightarrow$5 line intensities increase
by a factor of 1.7, 1.3, 1.1, and 1.1, respectively.  That is, the
lower $J$-lines are most affected, partly because of maser
action (in particular in the $J$=1$\rightarrow$0 line).  This means
that the SiO abundance estimates for the high mass-loss rate Miras are
particularly uncertain, and the line saturation makes things even
worse.

A velocity gradient may affect the SiO line intensities since it
allows the central pump photons to migrate further out in the CSE. We
have tested a velocity law of the form (appropriate for a dust-driven
wind, see \citet{habietal94})
\begin{equation}
\label{vlaw}
v(r) = \sqrt{v_{\rm i}^2 + \left( v_{\infty}^2-v_{\rm i}^2
            \right) \left( 1-\frac{r_{\rm i}}{r} \right)}\,,
\end{equation}
where $v_{\rm i}$ is the velocity at the inner radius, and $v_{\infty}$
the terminal velocity.  This produces a rather smooth increase in
velocity, and for $v_{\rm i}/v_{\infty}$\,=\,0.25 (which we have used)
90\% of the terminal velocity is reached at $r$\,=\,10$r_{\rm i}$ (for
low mass-loss rate objects, this is also the region which produces the
main part of the SiO radio line emission).  There is only an effect for the
low mass-loss rate object and the higher-$J$ lines.  For instance, the
$J$=6$\rightarrow$5 line intensity increases by about 10\%.  A
velocity gradient has though the effect that the lines become narrower
(Sect.~\ref{s:dynamics}).

Finally, the line intensity ratio $I$(SiO,$J$=2$\rightarrow$1)/
$I$(CO,$J$=1$\rightarrow$0) decreases with mass-loss rate: 0.79 for a
mass-loss rate of 10$^{-7}$\,M$_{\odot}$\,yr$^{-1}$, 0.33 for
10$^{-6}$\,M$_{\odot}$\,yr$^{-1}$, and 0.24 for
10$^{-5}$\,M$_{\odot}$\,yr$^{-1}$.  This result is in line with the
observational result presented in Fig.~\ref{f:sioco}, and suggests
that at least part of the trend is an excitation effect.

\section{CO modelling of the Miras}
\label{s:mirascomodel}

\begin{table*}
     \centering \caption[]{CO model results for the Mira sample}
     \label{t:COmodelresults} \[ \begin{tabular}{cccccclcccccc}
\hline
\noalign{\smallskip}
Source & $P$ & $L$\,$^1$ & $L_{\rm d}/L_{*}$ & $T_{*}$ & $T_{\rm d}$ & $D$ & $v_{\rm e}$ & $\dot{M}$ & $r_{\rm
p}$\,$^2$ & $h$ & $\chi^2_{\rm red}$ & N \\
       & [days] & [L$_\odot$] & & [K] & [K] & [pc] & [km s$^{-1}$] &
[10$^{-6}$\,M$_{\odot}$ yr$^{-1}$] & [10$^{16}$ cm] & & & \\
\noalign{\smallskip}
\hline
TX Cam      & 557 & {\phantom{1}}8400 & 0.26 & 1800 & 460 & 380 &             
18.5 & {\phantom{111}}7{\phantom{.0}} & {\phantom{1}}8.7 & 1.5 & 0.5 & 4 \\
R Cas       & 431 & {\phantom{1}}6500 & 0.06 & 2100 & 530 & 220 &
10.5 & {\phantom{111}}1.3             & {\phantom{1}}4.0 & 1.1 & 1.0 & 4 \\
R Hya       & 388 & {\phantom{1}}5800 & 0.03 & 2300 & 580 & 150 &
{\phantom{1}}7.0 & {\phantom{111}}0.3             & {\phantom{1}}2.0 & 0.5 &     & 1 \\
R Leo       & 313 & {\phantom{1}}4600 & 0.02 & 2100 & 570 & 130 &
{\phantom{1}}6.0 & {\phantom{111}}0.2             & {\phantom{1}}1.7 & 0.6 & 0.4 & 3 \\
GX Mon      & 527 & {\phantom{1}}8000 & 0.38 & 1500 & 380 & 540 &
18.7 &  {\phantom{11}}40{\phantom{.0}} &             24.0 & 0.5 & 2.4 & 4 \\
WX Psc      & 660 &             10000 & 0.85 & {\phantom{1}}920 & 400 & 600 &
19.3 &  {\phantom{11}}10{\phantom{.0}} &             11.4 & 0.4 & 6.6 & 4 \\
IK Tau      & 500 & {\phantom{1}}7500 & 0.46 & 1500 & 500 & 250 &
18.5 &  {\phantom{11}}30{\phantom{.0}} &             20.5 & 0.3 & 0.4 & 4 \\
IRC+10365   & 500 & {\phantom{1}}7500 & 0.26 & 1600 & 430 & 750 &
16.2 & {\phantom{11}}30{\phantom{.0}} &             23.8 & 0.5 & 2.2 & 2 \\
IRC$-$10529 & 680 &             10400 & 0.17 & 1000 & 410 & 270 &
12.0 & {\phantom{111}}2.5             & {\phantom{1}}5.8 & 0.1 & 1.6 & 4 \\
IRC$-$30398 & 575 & {\phantom{1}}8700 & 0.24 & 2000 & 480 & 390 &
16.0 & {\phantom{111}}6{\phantom{.0}} & {\phantom{1}}8.2 & 0.5 & 0.1 & 2 \\
IRC+40004   & 750 &             11500 & 0.44 & 1700 & 400 & 410 &
18.0 & {\phantom{111}}6{\phantom{.0}} & {\phantom{1}}8.6 & 0.5 & 2.4 & 2 \\
IRC+50137   & 629 & {\phantom{1}}9600 & 0.57 & 1300 & 310 & 410 &
17.0 & {\phantom{11}}10{\phantom{.0}} &             10.7 & 0.1 & 4.2 & 3 \\
\hline
\noalign{\smallskip}
\noalign{$^1$ Derived from a period--luminosity relation}
\noalign{$^2$ The CO photodissociation radius}
\end{tabular}
         \]
\end{table*}

In order to obtain mass-loss rates for the Mira sample we have modelled
the circumstellar CO radio line emission observed towards these stars
using the procedure described above and in 
\citet{schoolof01}.  The estimated mass-loss rates are given in
Table~\ref{t:COmodelresults}, rounded off to the number nearest to
1.0, 1.3, 1.5, 2.0, 2.5, 3, 4, 5, 6, or 8, i.e., these values are
separated by about 25\%.  The distribution of derived mass-loss rates
have a median value of 1.3$\times$10$^{-5}$\,M$_\odot$\,yr$^{-1}$.
Therefore, these Miras sample the high mass-loss rate end of AGB
stars.  Only two of them (\object{R~Hya} and \object{R~Leo}) have
low to intermediate mass-loss rates (a few times
10$^{-7}$\,M$_\odot$\,yr$^{-1}$).  $h$ was used as a free parameter in
the fit for those sources with more than two lines observed.  The
average value is 0.6, i.e., very similar to what
\citet{schoolof01} found for the more luminous stars in the
their carbon star sample.  We used $h$=0.5 for those stars observed in
only one or two lines.  The quality of the fits are given by the chi-square
statistic $\chi^2_{\rm red}$ (see Sect.~\ref{s:fitting} for the
definition).

A CO fractional abundance of 2$\times$10$^{-4}$ has been used
following the work of \cite{olofetal02} on the CO modelling of low to intermediate mass-loss
rate IRV/SRVs of M-type.  It is quite possible that, for the high
mass-loss rate stars involved here, the CO abundance is higher due to
a more efficient formation of CO at higher densities and lower
temperatures.  A higher CO abundance would lower somewhat the derived
mass-loss rates.

Among the Miras with the highest mass-loss rates there is a trend that the
model $J$=1$\rightarrow$0 line intensities are low for a model which
fits well the higher-$J$ lines.  The reason is that the CO lines reach the
saturation regime at about 10$^{-5}$\,M$_\odot$\,yr$^{-1}$, with the
higher-$J$ lines saturating first.
Therefore, we chose to put more weight on the high-$J$ lines in the
model fit.  The reported values for the mass-loss rates of these stars
are, in this context, therefore considered to be lower limits.  This
type of problem has also been encountered by 
\citet{kempetal03}.  For WX Psc, the only star in common with us, they derived a
mass-loss rate of 1.1$\times$10$^{-5}$\,M$_\odot$\,yr$^{-1}$ by
fitting the $J$=2$\rightarrow$1 line, and successively lower mass-loss
rates for the higher-$J$ lines reaching about
10$^{-6}$\,M$_\odot$\,yr$^{-1}$ by fitting the $J$=6$\rightarrow$5 and
$J$=7$\rightarrow$6 lines.  In this work a value of
1.1$\times$10$^{-5}$\,M$_\odot$\,yr$^{-1}$ is derived based on the
$J$=2$\rightarrow$1, $J$=3$\rightarrow$2, and $J$=4$\rightarrow$3
lines, but a fit to the $J$=1$\rightarrow$0 line requires a mass-loss
rate about a factor of three higher.  Kemper et al.  speculate that
variable mass loss and gradients in physical parameters (e.g., the
turbulent velocity width) may play a role.  To this we add that the
size of the CO envelope, which mainly affects low-$J$ lines, is
important.

The CO expansion velocities given in Table~\ref{t:COmodelresults} are
obtained in the model fits. Hence, they are somewhat more accurate
than a pure line profile fit, since for instance the effect of
turbulent broadening is taken into account.  The uncertainty is estimated to be
of the order $\pm$10\%.  The gas expansion velocities have a
distribution with a median value of 15.3\,km\,s$^{-1}$, while the
IRV/SRV sample has a median gas expansion velocity of
7.0\,km\,s$^{-1}$.  Again, only \object{R~Hya} and \object{R~Leo} have
low CO expansion velocities, below 10\,km\,s$^{-1}$.

\section{Size of the SiO envelope}
\label{s:siosize}

The results of the SiO line modelling will depend strongly on the
adopted sizes of the SiO envelopes.  Unfortunately, these are not
easily observationally determined nor theoretically estimated.  Early
work assumed that the whole CSE contributes to the observed SiO
thermal line emission (e.g., \citet{morretal79}).  The
mostly Gaussian-like SiO profiles found by  
\citet{bujaetal86, bujaetal89} towards O-rich CSEs
suggested that this is not the case.  The generally small size of the SiO
thermal line emitting region requires interferometric observations in
order to resolve it.  Results from SiO multi-line modelling and
interferometric data will be combined here to estimate the sizes of
the SiO envelopes.

\subsection{The SiO abundance distribution}
\label{s:abudistr}

Previous work strongly suggests that the SiO abundance in the CSE is
markedly lower than that in the stellar atmosphere. The decrease in the
SiO abundance with radius is very likely linked to two
different processes taking place in the CSE. Photodissociation due to
interstellar UV radiation is a well-known mechanism which reduces the
abundances of molecules in the extended CSE, but for SiO the depletion
onto grains closer to the star must also be taken into account.  We
outline here in a simplified way the effects of these processes [based
on the works by \citet{juramorr85, huggglas82}].  However, the theoretical results are not used in
our modelling, but they serve as a guide for the assumptions
and the interpretation.

Since the rate of evaporation is very large
for $T_{\rm gr}$\,$>$\,($T_{\rm bind}$/50) (where $T_{\rm gr}$
is the grain temperature, and $kT_{\rm bind}$ the binding energy of
the molecule onto grains), there is a critical radius, $r_{\rm 0}$,
such that for smaller radii there is effectively no condensation,
while for larger radii almost every molecule that sticks onto the
grain remains there.  The value of $r_{\rm 0}$ can be estimated from
the condition that the characteristic flow time, $r/v_{\rm e}(r)$, is
equal to the evaporation time $[R_{\rm evap}(T_{\rm gr})]^{-1}$.  A classical
evaporation theory has been used to obtain the rate for CO
\citep{lege83}, and the result is
\begin{equation}
\label{ro}
r_{\rm 0} = \frac{v_{\rm e}(r_{\rm 0})}{3\times10^{13}\,\exp
\left(-\frac{T_{\rm bind}}{T_{\rm gr}(r_{\rm 0})}\right)}\,\,{\rm cm},
\end{equation}
where $v_{\rm e}(r_{\rm 0})$ is given in cm\,s$^{-1}$.  While
different species have different coefficients in front of the
exponential, by far the most important term is the exponential.  The
rate of classical evaporation is generally so large that unless
$T_{\rm gr}$\,$\leq$\,$T_{\rm bind}/50$, condensation onto grains is
not important.  Therefore, in describing the condensation process,
only variations in $T_{\rm bind}$ for different substances are
considered and variations, among species, of the constant coefficient
in Eq.~(\ref{ro}) are ignored. $T_{\rm bind}$\,=\,29500\,K for
SiO \citep{legeetal85}, $T_{\rm
gr}(r)=T_{*}(R_{*}/2r)^{0.4}$ (appropriate for an optically thin dust
CSE), and $v_{\rm e}$\,=\,10\,km\,s$^{-1}$ results in a typical
condensation radius for our sources (with $L$\,=\,4000\,L$_{\odot}$
and $T_{\rm bb}$\,=2500\,K) of about 5$\times$10$^{14}$\,cm.

Using the formulation by
\citet{juramorr85}, the radial variation of the SiO abundance in a
CSE, taking into consideration the depletion of molecules onto dust
grains, is given by
\begin{equation}
\label{e:fcond}    
f_{\rm SiO}(r)= f_{\rm SiO}(r_0)\, \exp \left[ -r_{\rm scale}\left(\frac{1}{r_0} -\frac{1}{r}\right)\right],
\end{equation}
where $r_{\rm scale}$ is a scale length defined by
\begin{equation}
r_{\rm scale}=\frac{\alpha \dot{N}_{\rm d} \sigma_{\rm gr} v_{\rm dr}}{4\pi
v^2_{\rm e}}\,,
\end{equation}
where $\alpha$ is the sticking probability of SiO onto grains,
$\dot{N}_{\rm d}$ the dust mass-loss rate in terms of dust grain
number, $\sigma_{\rm gr}$ the grain cross section, and $v_{\rm dr}$
the drift velocity of the dust with respect to the gas, obtained
from the formula
\begin{equation}
v_{\rm dr} = \sqrt{ \frac{L v_{\rm e} Q_{\rm p,F}}{c \dot{M}}} \,.
\end{equation}
Thus, the abundance decreases due to condensation until it reaches the
terminal value
\begin{equation}
\label{e:finfty}    
f_{\rm SiO}(\infty) = f_{\rm SiO}(R_*)\, e^{-r_{\rm scale}/r_{0}}\,.
\end{equation}

The condensation efficiency depends strongly on the dust mass-loss
rate.  For instance, $\psi$\,=\,0.002 (appropriate for the average $h$-value
of the IRV/SRV sample), $a_{\rm gr}$\,=\,0.05\,$\mu$m, $\rho_{\rm
gr}$\,=\,2\,g\,cm$^{-3}$, $Q_{\rm p,F}$\,=\,0.03, $\alpha$=1, $L$\,=\,4000\,L$_{\odot}$,
$T_{\rm bb}$\,=2500\,K, and $v_{\rm e}$\,=\,10\,km\,s$^{-1}$ result in
$f_{\rm SiO}(\infty)/f_{\rm SiO}(R_*)$-values of 0.76, 0.42, and 0.07 for mass
loss rates of 10$^{-7}$\,M$_{\odot}$\,yr$^{-1}$,
10$^{-6}$\,M$_{\odot}$\,yr$^{-1}$, and
10$^{-5}$\,M$_{\odot}$\,yr$^{-1}$, respectively.  The corresponding
$f_{\rm SiO}(\infty)/f_{\rm SiO}(R_*)$-values for $\psi$\,=\,0.01 is 0.38, 0.013, and
10$^{-6}$.  Thus, we expect condensation to play only a minor role for
the low mass-loss rate objects, but its importance increases
drastically with the mass-loss rate.

The particular radius at
which the photodissociation becomes effective depends essentially on
the amount of dust in the envelope, which provides shielding against
the UV radiation, and the abundance of various molecular species if
the dissociation occurs in lines. \citet{huggglas82} describe the radial dependence of the
abundance  of a species of photospheric origin that is
shielded by dust (in the case of SiO,
shielding due to H$_2$O may be important but we ignore this here). 
Adopting this description in the case of SiO the result is
\begin{equation}
\frac{df_{\rm SiO}}{dr} = -\frac {G_{\rm 0,SiO}}{v_{\rm e}}\,\exp\left (
-\frac{d_{\rm SiO}}{r}\right)f_{\rm SiO},
\end{equation}
where $f_{\rm SiO}$ is the fractional abundance of SiO with respect to
H$_2$, $G_{\rm 0,SiO}$ the unshielded photodissociation rate of SiO,
and $d_{\rm SiO}$ the dust
shielding distance given by (see \citet{juramorr81})
\begin{equation}
d_{\rm SiO} = 1.4\,\frac{3(Q/a_{\rm gr})_{\rm SiO}}{4\rho_{\rm gr}}
                   \frac{\dot{M}_{\rm d}}{4\pi v_{\rm d}}
                   \:\propto\: \frac{h\dot{M}}{v_{\rm d}}\,,
\end{equation}
where $Q$ is the dust absorption efficiency, $\dot{M}_{\rm d}$ the 
dust mass-loss rate, and $v_{\rm d}$ the dust expansion
velocity given by $v_{\rm e} + v_{\rm dr}$.  The abundance decreases roughly 
exponentially with radius
and we adopt $f_{\rm SiO}$($r_{\rm p}$) = $f_{\rm SiO}$($R_{\ast}$)/e
to define the outer radius
$r_{\rm p}$.  It is obtained by solving the equation
\begin{equation}
\label{e:siophoto}
r_{\rm p}=\frac{v_{\rm e}/G_{\rm 0,SiO}}{E_2(d_{\rm SiO}/r_{\rm p})},
\end{equation}
where $E_2$(x) is the exponential integral.

Most likely the radial distribution of the SiO molecules is determined
by a combination of the condensation and photodissociation processes. 
Thus, one can imagine an initial SiO abundance determined by the
stellar atmosphere chemistry.  For low mass-loss rates, the abundance
decreases only slowly beyond the condensation radius until the
photodissociation effectively destroys all remaining SiO molecules. 
For high mass-loss rates, the abundance declines exponentially beyond
the condensation radius with an e-folding radius that can be estimated
from Eq.\,(\ref{e:fcond}),
 \begin{equation}
 r_{\rm c}=\left(\frac{1}{r_{\rm 0}}-\frac{1}{r_{\rm scale}}\right)^{-1}
 \end{equation}
(applicable only when $r_{\rm scale}$\,$>$\,$r_{\rm 0}$).  Using the same
parameters as above, except $\psi$\,=\,0.005 (appropriate for the
average $h$-value of the high mass-loss rate stars),
$L$\,=\,8000\,L$_{\odot}$, and $v_{\rm e}$\,=\,15\,km\,s$^{-1}$, the
result for 10$^{-5}$\,M$_{\odot}$\,yr$^{-1}$ is $r_{\rm
c}$\,=\,10$^{15}$\,cm, i.e., only about twice the condensation radius. 
An abundance decrease by a factor of a hundred is reached at about
4$\times$10$^{15}$\,cm, which is about a factor of five smaller than
the estimated SiO photodissociation radius for such an object.  Once
again, the results are sensitively dependent on the dust parameters,
e.g., $\psi$\,=\,0.002 results in $r_{\rm
c}$\,=\,2$\times$10$^{15}$\,cm, but the abundance (before
photodissociation) never decreases by more than a factor of five.

\subsection{The adopted SiO abundance distribution}

For the radial distribution of the SiO abundance in the CSEs we
adopt a Gaussian fall-off with increasing distance from the star,
\begin{equation}
\label{e:siodistrib}
f_{\rm SiO} = f_{\rm c}\,\,e^{-(r/r_{\rm e})^2},
\end{equation}
where $f_{\rm c}$ is the central abundance, and $r_{\rm e}$ the
e-folding distance.

This is a considerable simplification to the complicated SiO abundance
distribution.  However, as shown above, the expected distribution depends so
sensitively on the parameters adopted (in particular the dust mass
loss rate) that a more sophisticated approach is, for the moment, not
warranted.  We expect Eq.\,(\ref{e:siodistrib}) to be a reasonable
approximation to the SiO abundance distribution inside the
photodissociation radius for the low and intermediate mass-loss rate
objects. Equation\,(\ref{e:siodistrib}) is a reasonable
approximation for also the high mass-loss rate objects, but the size
is either determined by condensation (high $\psi$) or
photodissociation (low $\psi$).

We have checked whether the region within the condensation radius, with a
high SiO abundance, contributes substantially to the observed line
intensities.  For the model stars used in Sect.~\ref{s:dependence} it
is found that a high SiO abundance (5\,$\times$\,10$^{-5}$) inside the
condensation radius contributes by at most 20\% of the line
intensities from the rest of the SiO envelope.

\subsection{Results from SiO line modelling}
\label{s:siomodelresults}

The model code used in this work allow us to estimate SiO envelope sizes
provided that multi-line SiO observations are available.  The emission
from higher-$J$ lines comes very likely from the warmer inner regions
of the SiO envelope.  Therefore, the intensities of these lines can be
fitted by varying only the SiO abundance, i.e., they are rather insensitive
to the outer radius of the SiO envelope (see Table~\ref{t:pardep}). 
Once the SiO abundance has been found, the lower-$J$ lines can be used
as constraints to derive the size of the SiO envelopes, since their
emission is photodissociation limited (i.e., not excitation limited).

\begin{figure}
     \centerline{\psfig{file=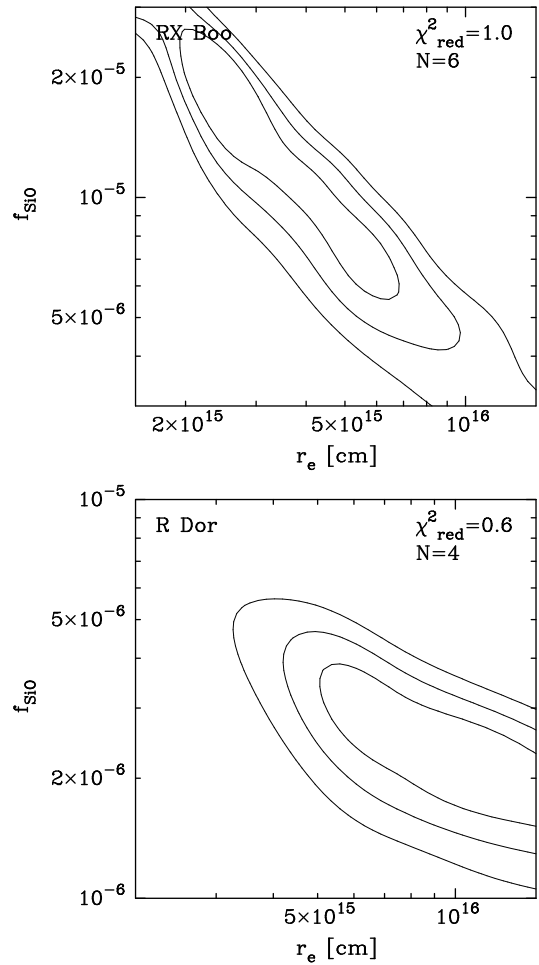,width=7.cm}}
     \caption{$\chi^2$ contours (at the 1, 2, and 3$\sigma$ levels) 
     in the SiO abundance and envelope size plane for
     two stars. In the case of \object{RX~Boo} there is a sufficient number of
     lines, including high-$J$ ones, to constrain the size of the SiO envelope.
     In the case of \object{R~Dor} the high-$J$ line data are missing and only a
     lower limit can be obtained (since we do not expect the SiO envelope to be
     larger than the CO envelope)}
     \label{f:chi2maps}
\end{figure}

It turns out that high-$J$ line data, e.g., $J$=8$\rightarrow$7, are 
required to constrain both the abundance and the size. These crucial high-$J$ 
line data were taken from \citet{biegetal00}. In the case of data
including only moderately high-$J$ lines, e.g., $J$=5$\rightarrow$4,
only a lower limit to the size can be obtained. This is illustrated in 
Fig.~\ref{f:chi2maps} where $\chi^2$ maps are given for two cases (see the
definition of the chi-square statistic below). In this
way, through the use of $\chi^2$ maps, we managed to estimate the SiO envelope 
sizes in 4 cases (\object{RX~Boo},
\object{R~Cas}, \object{IRC$-$10529}, \object{IRC+50137}), and obtain
lower limits to them in 7 cases (\object{TX~Cam}, \object{R~Crt}, \object{R~Dor},
\object{R~Leo}, \object{GX~Mon}, \object{L$^2$~Pup}, \object{IRC$-$30398}).

The resulting $r_{\rm e}$:s from the modelling are plotted as a 
function of the density measure $\dot{M}/v_{\rm
e}$, in Fig.~\ref{f:rerI}. We have here chosen to use
the lower limits to the SiO envelope sizes for all sources in
order to be consistent. The minimum least-square correlation between these SiO
envelope radii and the density measure is
\begin{equation}
\label{e:re}
\log r_{\rm e}\,=\,19.2 + 0.48\,\log\,\left(\frac{\dot M}{v_{\rm e}}\right)\,,
\end{equation}
(the correlation coefficient is 0.83) where $r_{\rm e}$ is given in cm, 
${\dot M}$ in M$_{\odot}$\,yr$^{-1}$, and $v_{\rm e}$ in km\,s$^{-1}$. 
For the rest of the sources the SiO envelope sizes could
not be derived through modelling.

\begin{figure}
     \centerline{\psfig{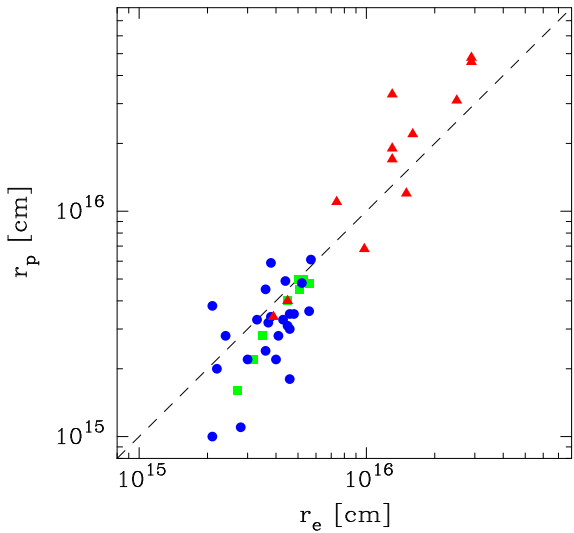}}
     \caption{SiO photodissociation radii (obtained using the unshielded
     photodissociation rate $G_0$\,=\,2.5\,$\times$\,10$^{-10}$\,s$^{-1}$) 
     versus the estimated sizes 
     of the SiO envelopes (IRV: square; SRV: circle; Mira: triangle). The dashed
     line shows the one-to-one relation}
     \label{f:rpre}
\end{figure}

We have checked our model results against those of the photodissocation
model. The photodissociation radii are estimated
from Eq.\,(\ref{e:siophoto}) assuming $Q$\,=\,1 \citep{suh00}
and using the appropriate $\dot{M}$- and $h$-values for each source.
A very good agreement with the estimated SiO envelope sizes
(for all sources with detected SiO lines),
from Eq.\,(\ref{e:re}), is obtained with an unshielded photodissociation
rate $G_{\rm 0,SiO}$\,=\,2.5\,$\times$\,10$^{-10}$\,s$^{-1}$ (the average
deviation is about 30\%),
see Fig.~\ref{f:rpre}. This value is lower by about a factor of two to
three than those reported by
\citet{vand88} and  \citet{taradalg90}, and higher by about a factor of two than that
reported by \citet{leteetal00}. The latter report
an uncertainty by (at least) a factor of two in their estimate. Thus, 
within the considerable uncertainties, our line modelling results are in excellent
agreement with those of the photodissociation model.
The $r_{\rm p}$:s for our sample are given in
Table~\ref{t:SiOmodelresults}.  On average, the photodissociation
radii of SiO are about a factor of 6 smaller than those of CO (the CO
results are given in cite{olofetal02}.

\subsection{Interferometry data}
\label{s:interferomdata}

We have also checked our modelling results by comparing with the interferometric 
SiO($J$=2$\rightarrow$1) data toward a number of O-rich
CSEs of \citet{lucaetal92}. They derived the sizes of the 
SiO line emitting region from direct
fits, assuming exponential source-brightness distributions, to the
visibility data. 
Their observations thus yielded the half-intensity angular radii of
the SiO($J$=2$\rightarrow$1) emitting regions.  \citet{sahabieg93} 
observed a smaller
sample of CSEs interferometrically, and claimed that the source
brightness distribution is rather of a power-law form (i.e.,
scale-free).  This would explain why Lucas et al.  derived essentially
the same angular sizes for most of the sources independent of their
distances.  To resolve this issue requires more detailed observations, and
we will only use the results of Lucas et al.  to compare with our modelling
results.

\begin{figure}
     \centerline{\psfig{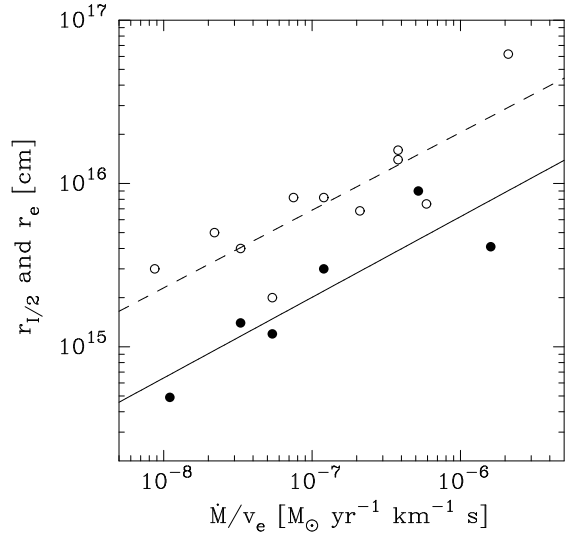}}
     \caption{ The sizes of the SiO envelopes estimated from the
     SiO line modelling are plotted versus a density measure (open circles).  
     The dashed line gives
     the relation between the SiO envelope size and the density measure
     given in Eq.\,(\ref{e:re}). Half intensity radii derived from interferometric
     SiO($J$=2$\rightarrow$1) observations are shown as solid
     circles.  The solid line is the fit to the data given in
     Eq.\,(\ref{e:ri}) }
     \label{f:rerI}
\end{figure}

We have six stars in common with \citet{lucaetal92} (RX~Boo,
R~Cas, W~Hya, R~Leo, WX~Psc, IK~Tau).  Fig~\ref{f:rerI} shows the
intensity radii as a function of the density measure $\dot{M}/v_{\rm
e}$, using our derived mass-loss rates, gas expansion velocities, and
distances.  The minimum least-square correlation between these intensity
radii and the density measure is
\begin{equation}
\label{e:ri}
\log r_{I/2}\,=\,18.8 + 0.49\,\log\,\left(\frac{\dot M}{v_{\rm e}}\right)\,,
\end{equation}
(the correlation coefficient is 0.88) where $r_{I/2}$ is given in cm, ${\dot M}$ in
M$_{\odot}$\,yr$^{-1}$, and $v_{\rm e}$ in km\,s$^{-1}$.

Thus, the scaling with the density measure of the intensity radii
is in perfect agreement with our modelling result for the 
envelope sizes. The estimated SiO envelope sizes that are required to model the
data are about three times larger than the SiO($J$=2$\rightarrow$1)
brightness region.  This may at first seem somewhat surprising, but a test
using the 10$^{-6}$\,M$_{\odot}$\,yr$^{-1}$ model star of
Sect.~\ref{s:dependence}, which has an SiO envelope radius of 1$\farcs$7,
shows that the resulting SiO($J$=2$\rightarrow$1) brightness
distribution has a half-intensity radius of 0$\farcs$4, i.e., about
four times smaller.

\section{Results of the SiO line model fits}

\subsection{The fitting procedure}
\label{s:fitting}

The radiative transfer analysis produces model brightness
distributions. These are convolved with the appropriate beams to allow
a direct comparison with the observed velocity-integrated line
intensities and to search for the best fit model. As observational constraints
we have used the data presented in this paper and the high-frequency data
obtained by \citet{biegetal00}. With the assumptions
made in the standard circumstellar model and the mass-loss rate and dust
properties derived from the modelling of circumstellar CO emission,
there remains only one free parameter, the SiO abundance [for all
stars $r_{\rm e}$ is taken from Eq.\,(\ref{e:re})].  The SiO abundance was
allowed to vary in steps of $\approx$10\% until the best-fit model was
found.  The quality of a particular model with respect to the
observational constraints can be quantified using the chi-square
statistic,
\begin{equation}
\label{chi2_eq}
\chi^2_{\rm red} = \frac{1}{\rm N-p}\,\sum^{\rm N}_{i=1}
\frac{[I_{\mathrm{mod},i} -
I_{\mathrm{obs},i})]^2}{\sigma^2_{i}},
\end{equation}
where $I$ is the total integrated line intensity, $\sigma_i$ the
uncertainty in observation $i$, p the number of free parameters (2 in
the cases of multi-line CO modelling, but only 1 
for the SiO line modelling, except in the cases discussed above where
also $r_{\rm e}$ was a free parameter), and the summation is done over
all independent observations N. The errors in the observed intensities
are always larger than the calibration uncertainty of $\approx$20\%. 
We have chosen to adopt $\sigma_i$\,=\,0.2$I_{\mathrm{obs},i}$ to put
equal weight on all lines, irrespective of the S/N-ratio.  The final
chi-square values for stars observed in more than one transition are
given in Table~\ref{t:SiOmodelresults}. They are, in many cases, rather
large suggesting that our circumstellar model may not be entirely appropriate
for the modelling of the SiO radio line emission, see Sect.~\ref{s:discussion}.
The line profiles were not
used to discriminate between models, but differences between model and
observed line profiles are discussed in Sects~\ref{s:dynamics} and
\ref{s:discussion}.

\subsection{The accuracy of the estimated abundances}

We will here try to estimate the uncertainty in the derived SiO
abundances. The uncertainties due to the adopted circumstellar model
are ignored since these are very difficult to estimate, and
focus is put on those introduced by the adopted parameters (see
Sect.~\ref{s:dependence}).  We start by considering the IRV/SRVs.  The
results depend crucially on the validity of Eq.\,(\ref{e:re}).  A
change by $-$50\% and +100\% in the size of the SiO envelope results
in a variation of the $J$=2$\rightarrow$1 line intensity by about
$\pm$50\%, and therefore an equal uncertainty in the abundance.  The
product of $f_{\rm SiO}$ and $\dot{M}$ is essentially constant for a
best fit model.  It is estimated that the mass-loss rate is uncertain
by at least a factor of two (due to the modelling).  An uncertainty in
the distance has only a minor effect on the abundance (the change in
mass-loss rate compensates for the change in distance).  The
dependence on the luminosity is moderate.  We therefore estimate that,
within the adopted circumstellar model, the derived SiO abundances are
uncertain by at least a factor of three for those sources with
multi-line observations.  The uncertainty increases to a factor of
five when only one transition is observed.

For the high mass-loss rate (i.e.,
$\gtrsim$\,5$\times$10$^{-6}$\,M$_{\odot}$\,yr$^{-1}$) Miras the situation is even
worse.  The radiation from these stars are significantly converted
into longer-wavelength dust radiation, which has been taken care of
only crudely by using two central blackbodies.  Tests show that the
resulting SiO line intensities are sensitive to the structure of the
radiation sources, Sect~\ref{s:dependence}.  In addition, the SiO
lines are rather saturated and hence the line intensities are, at
least partly, insensitive to the abundance.  Therefore, it is
estimated that for these objects the SiO abundance is uncertain
by a factor of five (in all cases information on three, or more, lines is
available), but note that any reasonable change in the
radiation field structure will systematically lower the abundance
required to fit the data.

\begin{table*}
     \centering
     \caption[]{Source parameters and SiO model results}
      \label{t:SiOmodelresults}
    \[
	\begin{tabular}{llccccccccrc}
\hline
\noalign{\smallskip}
Source & Var.  & $P$ & $D$ & $v_{\rm e}$(SiO) & $v_{\rm e}$(CO)
 & $\dot{M}$ & $r_{\rm p}$ & $r_{\rm e}$ & $f_{\rm SiO}$ & $\chi^2_{\rm red}$ & N \\
 & type & [days] & [pc] & [km s$^{-1}$] & [km s$^{-1}$]
&[10$^{-7}$\,M$_{\odot}$ yr$^{-1}$] & [10$^{15}$ cm] & [10$^{15}$ cm] & [10$^{-6}$] & & \\
\hline
RS And   & SRa & 136 & 290$^1$            &  {\phantom{1}}4.4 &
{\phantom{1}}4.4  & {\phantom{11}}1.5               & {\phantom{1}}2.2 &
{\phantom{1}}4.0 &              16{\phantom{.0}} &     & 1 \\
UX And   & SRb & 400 & 280$^1$            &              12.8 &
12.8  & {\phantom{11}}4{\phantom{.0}}   & {\phantom{1}}5.9 &
{\phantom{1}}3.8 &              12{\phantom{.0}} &     & 1 \\
$\theta$ Aps & SRb & 119 & 110$^1$        &  {\phantom{1}}4.0 &
{\phantom{1}}4.5  & {\phantom{11}}0.4               & {\phantom{1}}2.0 &
{\phantom{1}}2.2 &              14{\phantom{.0}} & 0.3  & 3\\
TZ Aql   & Lb  &     & 470$^1$            &  {\phantom{1}}4.8 &
{\phantom{1}}4.8  & {\phantom{11}}1{\phantom{.0}}   & {\phantom{1}}2.2 &
{\phantom{1}}3.2 &              15{\phantom{.0}} & 8.1 & 2 \\
SV Aqr   & Lb  &     & 470$^1$            &  {\phantom{1}}8.0 &
{\phantom{1}}8.0  & {\phantom{11}}3{\phantom{.0}}   & {\phantom{1}}4.0 &
{\phantom{1}}4.5  &              34{\phantom{.0}} & 2.1 & 2 \\
T Ari    & SRa & 317 & 310$^1$            &  {\phantom{1}}2.4 &
{\phantom{1}}2.4  & {\phantom{11}}0.4               & {\phantom{1}}1.1 &
{\phantom{1}}2.8 &  {\phantom{1}}5.2 & 2.9 & 2 \\
RX Boo   & SRb & 340 & 110$^1$            &  {\phantom{1}}7.8 &
{\phantom{1}}9.3  & {\phantom{11}}5{\phantom{.0}}   & {\phantom{1}}4.8 &
{\phantom{1}}5.2 &  {\phantom{1}}8.0 & 1.2 & 6 \\
RV Cam   & SRb & 101 & 350$^1$            &  {\phantom{1}}5.8 &
{\phantom{1}}5.8  & {\phantom{11}}2.5               & {\phantom{1}}3.0 &
{\phantom{1}}4.6 &  {\phantom{1}}4.5 &     & 1 \\
TX Cam   &  M  & 557 & 380{\phantom{$^1$}}&              16.0 &
18.5  & {\phantom{1}}60{\phantom{.0}}   & 33{\phantom{.0}}             &
13{\phantom{.0}} &   {\phantom{1}}5.5 & 2.0    & 3 \\
R Cas    &  M  & 431 & 220{\phantom{$^1$}}&  {\phantom{1}}7.0 &
10.5  & {\phantom{1}}13{\phantom{.0}}   & 11{\phantom{.0}} &
{\phantom{1}}7.4 &      {\phantom{1}}7.0 &  0.8   & 3 \\
UY Cet   & SRb & 440 & 300$^1$            &  {\phantom{1}}6.0 &
{\phantom{1}}6.0  & {\phantom{11}}2.5	            & {\phantom{1}}3.1 &
{\phantom{1}}4.5 &  {\phantom{1}}6.0 & 3.0 & 3 \\
CW Cnc   & Lb  &     & 280$^1$            &  {\phantom{1}}7.0 &
{\phantom{1}}8.5  & {\phantom{11}}5{\phantom{.0}}   & {\phantom{1}}4.5 &
{\phantom{1}}5.1 &  {\phantom{1}}2.7 &     & 1 \\
R Crt    & SRb & 160 & 170$^1$            &              10.6 &
10.6  & {\phantom{11}}8{\phantom{.0}}   & {\phantom{1}}6.1 &
{\phantom{1}}5.7 &  {\phantom{1}}6.0 & 2.5 & 3 \\
R Dor    & SRb & 338 & {\phantom{1}}45$^1$             &  {\phantom{1}}5.0 &
{\phantom{1}}6.0  & {\phantom{11}}1.3	            & {\phantom{1}}3.3 &
{\phantom{1}}3.3 &  {\phantom{1}}5.0 & 2.4 & 4 \\
AH Dra   & SRb & 158    & 340$^1$            &              {\phantom{1}}6.4 &
{\phantom{1}}6.4     & {\phantom{11}}0.8          & {\phantom{1}}2.0 &
{\phantom{1}}2.6 &        17{\phantom{.0}} &     & 1 \\
CS Dra   & Lb  &     & 370$^1$            &              11.6 &
11.6     & {\phantom{11}}6{\phantom{.0}}   & {\phantom{1}}5.0 &
{\phantom{1}}5.0 &  {\phantom{1}}2.7 &     & 1 \\
S Dra    & SRb & 136 & 270$^1$            &  {\phantom{1}}9.6 &
{\phantom{1}}9.6  & {\phantom{11}}4{\phantom{.0}}   & {\phantom{1}}4.9 &
{\phantom{1}}4.4 &  {\phantom{1}}7.0 &  4.3   & 2 \\
SZ Dra   & Lb  &     & 510$^1$            &  {\phantom{1}}9.6 &
{\phantom{1}}9.6  & {\phantom{11}}6{\phantom{.0}}   & {\phantom{1}}5.0 &
{\phantom{1}}5.3 &  {\phantom{1}}1.8 &     & 1 \\
TY Dra   & Lb  &     & 430$^1$            &  {\phantom{1}}9.0 &
{\phantom{1}}9.0  & {\phantom{11}}6{\phantom{.0}}   & {\phantom{1}}4.8 &
{\phantom{1}}5.6 &  10{\phantom{.0}} &  0.3   & 2 \\
R Hya    &  M  & 388 & 150{\phantom{$^1$}}&  {\phantom{1}}4.5 &
{\phantom{1}}7.0  & {\phantom{11}}3{\phantom{.0}}   & {\phantom{1}}4.0 &
{\phantom{1}}4.5 &         {\phantom{1}}7.0 &  4.2   & 2 \\
W Hya    & SRa & 361 & {\phantom{1}}65$^1$             &  {\phantom{1}}6.5 &
{\phantom{1}}6.5  & {\phantom{11}}0.8	            & {\phantom{1}}2.8 &
{\phantom{1}}2.4  & 15{\phantom{.0}} & 4.5 & 7 \\
R Leo    &  M  & 313 & 130{\phantom{$^1$}}&  {\phantom{1}}6.0 &
{\phantom{1}}6.0  & {\phantom{11}}2.0 	            & {\phantom{1}}3.9 &
{\phantom{1}}5.5 &              13{\phantom{.0}} & 2.5 & 3 \\
U Men    & SRa & 407 & 320$^1$            &  {\phantom{1}}7.2 &
{\phantom{1}}7.2  & {\phantom{11}}2.0	            & {\phantom{1}}3.4 &
{\phantom{1}}3.8 &  {\phantom{1}}5.8 & 3.5 & 2 \\
T Mic    & SRb & 347 & 130$^1$            &  {\phantom{1}}4.8 &
{\phantom{1}}4.8  & {\phantom{11}}0.8	            & {\phantom{1}}2.2 &
{\phantom{1}}3.0 &  {\phantom{1}}5.3 & 1.2 & 2 \\
GX Mon   &  M  & 527 & 540{\phantom{$^1$}}&              18.7 &
18.7  & 400{\phantom{.0}}               & 48{\phantom{.0}}    &
29{\phantom{.0}} & {\phantom{1}}0.8 &  4.0   & 5 \\
S Pav    & SRa & 381 & 150$^1$            &  {\phantom{1}}4.8 &
{\phantom{1}}9.0  & {\phantom{11}}0.8	            & {\phantom{1}}3.8 &
{\phantom{1}}2.1 &  {\phantom{1}}2.6 & 0.3 & 2 \\
SV Peg   & SRb & 145 & 190$^1$            &  {\phantom{1}}6.3 &
{\phantom{1}}7.5  & {\phantom{11}}3{\phantom{.0}}   & {\phantom{1}}3.5 &
{\phantom{1}}6.5 &  {\phantom{1}}5.1 &     & 1 \\
TW Peg   & SRb & 929 & 200$^1$            &  {\phantom{1}}9.5 &
{\phantom{1}}9.5  & {\phantom{11}}2.5	            & {\phantom{1}}4.5 &
{\phantom{1}}3.6 &  {\phantom{1}}2.4 &     & 1 \\
WX Psc   &  M  & 660 & 600{\phantom{$^1$}}&              19.3 &
19.3  & 110{\phantom{.0}}               & 22{\phantom{.0}}             &
16{\phantom{.0}} &              {\phantom{1}}6.0 &  3.1   & 4 \\
L$^2$ Pup& SRb & 141 & {\phantom{1}}85$^1$             &  {\phantom{1}}2.3 &
{\phantom{1}}2.3  & {\phantom{11}}0.2	            & {\phantom{1}}1.0 &
{\phantom{1}}2.1 &  14{\phantom{.0}} & 2.1 & 3 \\
Y Scl    & SRb &     & 330$^1$            &  {\phantom{1}}5.2 &
{\phantom{1}}5.2  & {\phantom{11}}1.3	            & {\phantom{1}}2.4 &
{\phantom{1}}3.6 &  {\phantom{1}}5.0 & 0.7 & 2 \\
V1943 Sgr& Lb  &     & 150$^1$            &  {\phantom{1}}4.6 &
{\phantom{1}}5.4  & {\phantom{11}}1.3	            & {\phantom{1}}2.8 &
{\phantom{1}}3.5 &  {\phantom{1}}7.3 & 1.3 & 2 \\
IK Tau   &  M  & 500 & 250{\phantom{$^1$}}&              17.5 &
18.5  & 300{\phantom{.0}}               &   31{\phantom{.0}}   &
25{\phantom{.0}} &  {\phantom{1}}0.4 & 2.8 & 4 \\
V Tel    & SRb & 125 & 290$^1$            &  {\phantom{1}}6.8 &
{\phantom{1}}6.8  & {\phantom{11}}2.0	            & {\phantom{1}}3.2 &
{\phantom{1}}3.7  &  {\phantom{1}}5.0 & 12.0 & 2 \\
Y Tel & Lb &   & 340$^1$            &  {\phantom{1}}3.5 &
{\phantom{1}}3.5  & {\phantom{11}}5{\phantom{.0}}     & {\phantom{1}}1.6 &
{\phantom{1}}2.7 &              54{\phantom{.0}} & 0.1  & 2 \\
Y UMa    & SRb & 168 & 220$^1$            &  {\phantom{1}}4.8 &
{\phantom{1}}4.8  & {\phantom{11}}1.5	            & {\phantom{1}}2.8 &
{\phantom{1}}4.1 &              12{\phantom{.0}} &     & 1 \\
SU Vel   & SRb & 150 & 250$^1$            &  {\phantom{1}}5.5 &
{\phantom{1}}5.5  & {\phantom{11}}2.0	            & {\phantom{1}}3.3 &
{\phantom{1}}4.3 &  {\phantom{1}}2.7 & 5.3 & 2 \\
BK Vir   & SRb & 150 & 190$^1$            &  {\phantom{1}}4.0 &
{\phantom{1}}4.0  & {\phantom{11}}1.5	            & {\phantom{1}}1.8 &
{\phantom{1}}4.6 &  {\phantom{1}}2.3 &     & 1 \\
RT Vir   & SRb & 155 & 170$^1$            &  {\phantom{1}}6.2 &
{\phantom{1}}7.8  & {\phantom{11}}5{\phantom{.0}}   & {\phantom{1}}3.6 &
{\phantom{1}}5.6 &              13{\phantom{.0}} & 1.3 & 2 \\
SW Vir   & SRb & 150 & 120$^1$            &  {\phantom{1}}7.5 &
{\phantom{1}}7.5  & {\phantom{11}}4{\phantom{.0}}   & {\phantom{1}}3.5 &
{\phantom{1}}4.8 &  {\phantom{1}}3.5 &  9.1   & 4 \\
IRC+10365&  M  & 500 & 750{\phantom{$^1$}}&              16.2 &
16.2  & 300{\phantom{.0}}               &     46{\phantom{.0}}     &
29{\phantom{.0}} &            {\phantom{1}}4.0 &  3.9   & 3 \\
IRC$-$10529&  M  & 680 & 270{\phantom{$^1$}}&              12.0 &
12.0  & {\phantom{1}}25{\phantom{.0}}   & {\phantom{1}}6.8  &
10{\phantom{.0}} &  {\phantom{1}}1.1 &  0.1   & 3 \\
IRC$-$30398&  M  & 575 & 390{\phantom{$^1$}}&              16.0 &
         16.0  & {\phantom{1}}60{\phantom{.0}}&    17{\phantom{.0}}  &
13{\phantom{.0}} &  {\phantom{1}}0.3 &  10.0   & 3 \\
IRC+40004&  M  & 750 & 410{\phantom{$^1$}}&              18.0 &
18.0  & {\phantom{1}}60{\phantom{.0}}   &      19{\phantom{.0}} &
13{\phantom{.0}} &  {\phantom{1}}0.2 &   8.7   & 3 \\
IRC+50137&  M  & 629 & 410{\phantom{$^1$}}&              17.0 &
17.0  & 100{\phantom{.0}}               &     12{\phantom{.0}}  & 
15{\phantom{.0}} &  {\phantom{1}}0.5 &   2.0   & 3 \\
\hline
\noalign{\smallskip}
\noalign{$^1$ Distance derived assuming a luminosity of 4000\,L$_{\odot}$}
\end{tabular}
         \]
\end{table*}

\begin{figure}
     \centerline{\psfig{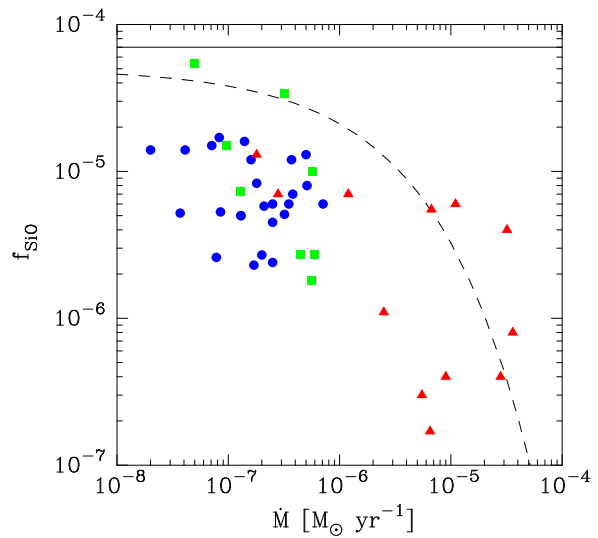}} 
     \caption{SiO fractional abundances versus the mass-loss rate
     (IRV:square, SRV:circle, Mira:triangle).  The horizontal line marks
     the maximum abundance allowed by solar abundances.  The dashed line
     shows the expected $f(\infty)$ (scaled to 5$\times$10$^{-5}$, roughly
     the expected abundance from stellar equilibrium chemistry, at
     very low mass-loss rates) for the parameters given in
     Sect.~\ref{s:discussion}}
     \label{abundancemassloss}
\end{figure}

\subsection{Abundances}
\label{s:abundances}

It can be assumed that the stars in our samples have silicon abundances close
to the solar value, Si/H\,=\,3.6$\times$10$^{-5}$ 
\citep{andegrev89}.  If Si is fully associated with O as SiO, and all H
is in H$_2$, the maximum SiO fractional abundance is
7$\times$10$^{-5}$.  Detailed calculations on stellar atmosphere
equilibrium chemistry give abundances in the vicinity of this for M-stars, 
about 4$\times$10$^{-5}$ \citep{duaretal99}.  Duari et al. 
also show that the SiO abundance is not affected by atmospheric shocks in the
case of M-stars.

The derived SiO abundances are given in Table~\ref{t:SiOmodelresults}.
The distribution for the IRV/SRV sample has a median value of
6$\times$10$^{-6}$, and a minimum of 2$\times$10$^{-6}$ and a maximum
of 5$\times$10$^{-5}$. For the IRVs and SRVs the median results are
9$\times$10$^{-6}$ and 6$\times$10$^{-6}$, respectively.  This is
almost a factor of ten lower than expected from theory. 
Figure~\ref{abundancemassloss} shows the SiO abundance as a function of
the mass-loss rate.  In addition to the abundances being low, there
is also a trend in the sense that both the upper and the lower `envelope'
of the abundances decrease with increasing mass-loss rate.

\begin{figure*}[t]    
\centerline{{\includegraphics[width=15.0cm]{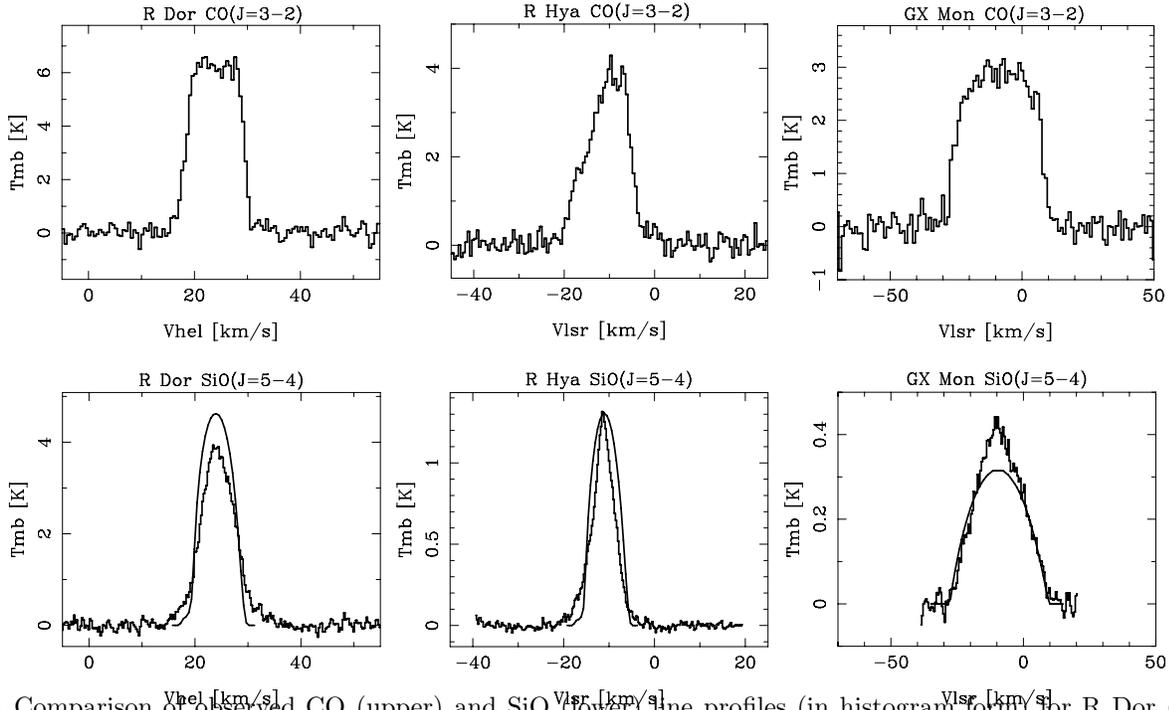}}}
        \caption{Comparison of observed CO (upper) and SiO (lower) line profiles (in
        histogram form) for \object{R~Dor} (left) and \object{R~Hya} (middle) and \object{GX~Mon}
        (right). The corresponding best-fit (i.e., to all observed line intensities) model 
        SiO lines are also shown as solid lines}
        \label{f:siocospectra}
\end{figure*}

The low mass-loss rate Miras follow the trend of the IRV/SRVs, 
and for the high mass-loss rate ($\dot{M} > 5 \times 10^{-6}$\,M$_{\odot}$\,yr$^{-1}$)
Miras we find a substantially lower abundance, a median below 10$^{-6}$.
Thus, the inclusion of the Miras shows that the trend of
decreasing SiO abundance with increasing mass-loss rate continues
towards high mass-loss rates.  This is further discussed in
Sect.~\ref{s:discussion} where an interpretation in terms of increased
adsorption of SiO onto dust grains the higher the mass-loss rate is
advocated.

The spread in abundance, at a given mass-loss rate, is
substantial, but it is within the (considerable) uncertainties, except
possibly for the high mass-loss rate Miras, for which there seem
to be a division into a low
abundance group (on average 4$\times$10$^{-7}$) and a high abundance
group (on average 5$\times$10$^{-6}$), while .  This division into two 
well-separated groups is
peculiar, but within the circumstellar model used here this conclusion
appears inescapable.  One can argue that the modelling of the high 
mass-loss rate Mira SiO line emission is particularly difficult, but we
find no reason why errors in the model should affect stars with essentially
similar properties ($L$, $\dot{M}$, $v_{\rm e}$) so differently.

\subsection{CSE dynamics}
\label{s:dynamics}

The observed SiO line profiles are used in the modelling to derive the gas
expansion velocities in the regions of the CSEs where the observed SiO
line emission stems from.  A comparison of these values with the gas
expansion velocities derived from the modelling of circumstellar CO line
emission is indeed a direct probe of the CSE dynamics since the
extents of the SiO and CO line emitting regions are very different.

The SiO and CO radio line profiles are clearly different, although
this conclusion is mainly based on the limited number of sources where
the S/N-ratio of the data are high enough for both species.  In
Table~\ref{t:SiOmodelresults} different values for the gas
expansion velocity estimated from the SiO and the CO data are reported in the 11
cases where these are regarded as significantly different.  In all cases
the SiO velocities are smaller than
those obtained from the CO data.  Indeed, the SiO line profiles are
narrower in the sense that the main fraction of the emission comes
from a velocity range narrower than twice the expansion velocity
determined from the CO data.  On the other hand, the SiO line profiles
have weak wings so that the total velocity width of its emission is
very similar to that of the CO emission.  This is illustrated in
Fig.~\ref{f:siocospectra}, where we also show the corresponding best-fit (i.e., to
all observed line intensities) model SiO lines. It is clear that the model line
profiles do not provide perfect fits to the observed line profiles, but
they show that for the lower mass-loss rate objects the SiO line profiles
are strongly affected by selfabsorption on the blue-shifted side. This 
explain partly why the SiO lines are narrower than the CO lines.
The remaining discrepancy  is interpreted as due to the influence of gas acceleration
in the region which produces a significant fraction of the SiO line
emission, as suggested already by \cite{bujaetal86}.  
This interpretation is quantitatively corroborated by our
modelling results when a velocity gradient is included, see Sect.~\ref{s:dependence}.  
The extent of the effect is though uncertain.  \citet{biegetal00}, by
comparing high-$J$ SiO lines with CO line data, concluded that the SiO
lines are formed predominantly in the part of the CSE where the gas
velocity exceeds 90\% of the terminal velocity.  We suspect
that the discrepancy between the widths of the SiO and CO lines
decreases with the mass-loss rate of the object. In addition, we 
find that for at least some of the high-mass-loss-rate sources the 
higher-$J$ SiO lines become essentially
triangular, see \object{GX~Mon} in Fig.~\ref{f:siocospectra}. 
The model does a fairly good job 
in reproducing these SiO line profiles, except that the model lines
are less sharply peaked. A high sensitivity,
multi-line study combined with interferometric observations are
required to fully tackle this problem.

\begin{figure}
     \centerline{\psfig{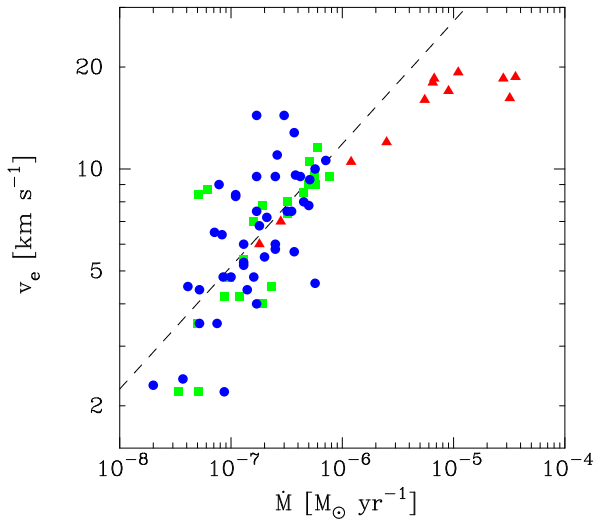}} 
     \caption{The derived CO gas expansion velocities as a function of the mass-loss
     rates for the IRV (squares), SRV (circles), and Mira (triangles) samples. The
     dashed line shows the correlation found for the IRV/SRV sample (see text)}
     \label{f:vexpml}
\end{figure}

In this connection we also present Fig.~\ref{f:vexpml} which shows the
gas expansion velocity (determined from CO line modelling) as a function of mass
loss rate for the IRV/SRV and Mira samples.  This is an extension of
the result of \citet{olofetal02}, and it shows that
low to intermediate mass-loss rate winds have a scaling of $v_{\rm e}
\propto \dot{M}^{0.36}$, and that this gradually goes over into a wind
of close to 20\,km\,s$^{-1}$, for higher mass-loss rates. This is as 
expected for a dust-driven wind \citep{elitivez01}.

\subsection{Peculiar sources}
\label{s:pecsources}

To single out peculiar sources is a highly subjective process, and it
also depends strongly on the S/N-ratio of the data (at high enough
S/N-ratio probably most sources show a deviation from the
expected). Here, a few sources in the IRV/SRV
sample which qualify as peculiar or for which we have problems in the
SiO line modelling are discussed.

\citet{kersolof99} found four objects in their
sample of circumstellar CO radio line emission, which clearly show
double-component line profiles, a narrow feature centred on a broad plateau
(\object{EP~Aqr}, \object{RV~Boo}, \object{X~Her}, and
\object{SV~Psc}), all of them SRVs.  
\citet{olofetal02} determined mass-loss rates and gas expansion
velocities by simply decomposing the emission into two components and
assuming that the emissions are additive.  They found that the mass
loss rates are higher for the broader component by, on average, an
order of magnitude.  The gas expansion velocities derived from the
narrow components ($\approx$1.5 km\,s$^{-1}$) put to question an
interpretation in the form of a spherical outflow.  The origin of such
a line profile is still not clear (see 
\citet{olofetal02} for a discussion on this issue).  These four sources
are also included in our SiO sample, and the spectra are shown in
Figs.~\ref{f:siosr1} and \ref{f:siosr2}.

Towards \object{EP~Aqr} there is no sign of the narrow feature in the
SiO $J$=2$\rightarrow$1 and $J$=3$\rightarrow$2 lines, only the broad
feature is clearly present.  This suggests that the broad feature
originates in a `normal' CSE, while the narrow feature may have a
different origin.  We note though that the SiO line profile of the
broad component deviates somewhat from a smooth symmetric profile. 
\object{SV~Psc} is very similar to \object{EP~Aqr} in CO in the sense
that the narrow feature is very much narrower than the broad feature. 
Unfortunately, the \object{SV~Psc} SiO data are of low quality, but
both components appear to be present.  In the cases of \object{RV~Boo}
and \object{X~Her} the CO and SiO line profiles are very similar, and
the widths of the narrow components are about half of those of the
broad ones.  The SiO abundances of both components have been obtained,
assuming that the emissions are additive.  The results are given in
Table~\ref{t:doublecomponents}.  For all sources, and for both
components, the results appear normal.

\begin{figure*}
\sidecaption
     \includegraphics[width=12cm]{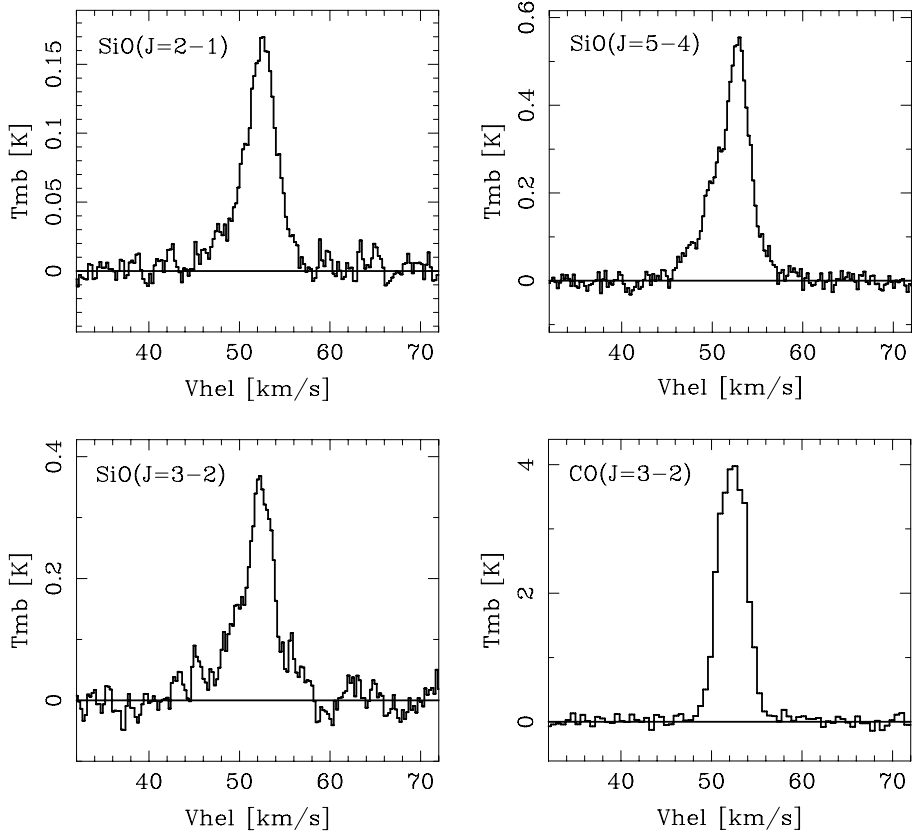}
     \caption{\object{L$^2$~Pup} SiO line spectra and a CO($J$=3$\rightarrow$2)
     spectrum (from \citet{olofetal02})}
     \label{f:l2pupdata}
\end{figure*}

\begin{table*}
     \centering
     \caption[]{Source parameters and model results for those objects
     with double component line profiles}
      \label{t:doublecomponents}
    \[
     \begin{tabular}{llrrlccccccc}
     \hline
     \noalign{\smallskip}
Source & Var.  & $P${\phantom{00}} & $D$ & comp.  & $\dot{M}$ & $v_{\rm e}$(SiO)
& $v_{\rm e}$(CO) & $r_{\rm e}$ & $f_{\rm SiO}$ & N \\
& type & [days] & [pc] & & [10$^{-7}$\,M$_{\odot}$\,yr$^{-1}$] &
[km\,s$^{-1}$] & [km\,s$^{-1}$] & [10$^{15}$ cm] &  [10$^{-6}$] &\\
\hline
EP Aqr   & SRb & {\phantom{1}}55  & 140  & broad  & 5.0 & 7.8 & 9.2  & 5.2 & {\phantom{1}}3.8 &  2\\
RV Boo   & SRb &              137 & 280  & broad  & 2.0 & 6.8 & 7.0  & 3.2 & {\phantom{1}}6.0 &  2 \\
         &      &                 &      & narrow & 0.3 & 3.0 & 2.3  & 2.6 & {\phantom{1}}7.0 &  1 \\
X Her    & SRb  & {\phantom{1}}95  & 140 & broad  & 1.5 & 6.5 & 6.5  & 3.4 &             12{\phantom{.0}} &  1 \\
         &      &                 &      & narrow & 0.4 & 2.5 & 2.2  & 3.1 & {\phantom{1}}4.0 &  1 \\
SV Psc   & SRb &              102 & 380  & broad  & 3.0 & 8.6 & 9.5  & 4.0 &             10{\phantom{.0}} &  3 \\
		 &     &                  &      & narrow & 0.4 & 1.6 & 2.2  & 3.2 & {\phantom{1}}6.0 &  1\\ 
\hline
\end{tabular}
         \]
\end{table*}

\object{L$^2$~Pup} was singled out in 
\citet{olofetal02} as a low mass-loss rate
(2$\times$10$^{-8}$\,M$_{\odot}$\,yr$^{-1}$), low gas expansion
velocity (2.1\,km\,s$^{-1}$) object.  This star has been recently
discussed also by \citet{juraetal02} and 
\citet{wintetal02}.  In the latter paper comparisons are made with wind
models, and it is concluded that stars with the mass-loss properties
of \object{L$^2$~Pup} can be understood in terms of a pulsationally
driven wind, where dust plays no dynamic role. Our SiO line profiles 
resemble to some extent those of CO in the sense that the narrow
feature is also present.  However, the SiO lines clearly show broad
line wings, Fig.~\ref{f:l2pupdata}.  The full
velocity width of these lines are $\approx$\,12\,km\,s$^{-1}$, i.e.,
larger than the CO line width, but narrower than the
SiO($v$=1, $J$=2$\rightarrow$1) maser line width of
$\approx$\,20\,km\,s$^{-1}$ measured by 
\citet{wintetal02}.  In addition, the narrow feature, which appears
narrower in the SiO lines than in the CO lines (Fig.~\ref{f:l2pupdata}),
is not exactly centered on the broad component, its center lies at 
$v_{\rm hel}$\,=\,52.8\,km\,s$^{-1}$ as opposed to 51.4\,km\,s$^{-1}$ for the latter. 
This suggests a rather complicated dynamics in the
inner part of the CSE, but high-quality data, also in higher-$J$ SiO
lines, are required before progress can be made.

\object{W~Hya} is one of the sources for which we have the highest
quality data.  It is also one of the sources with the poorest best-fit model.
A much better fit is obtained by increasing the size of the SiO
envelope to $r_{\rm e}$=6$\times$10$^{15}$\,cm (and  $f_{\rm
SiO}$=8$\times$10$^{-6}$ as determined from the high-$J$ lines), i.e.,
almost a factor of three higher than that obtained from Eq.({\ref{e:re}). 
Considering the uncertainties this is of no major concern.  However,
it is worth recalling that \citet{olofetal02}
derived a (molecular hydrogen) mass-loss rate of
7$\times$10$^{-8}$\,M$_{\odot}$\,yr$^{-1}$ from CO data [this result
has been confirmed by including CO $J$=1$\rightarrow$0 and 2$\rightarrow$1
IRAM 30\,m data
\citep{bujaetal89, cernetal97a}, CO $J$=2$\rightarrow$1,
3$\rightarrow$2, and 4$\rightarrow$3 JCMT archive data, and the CO ISO results
of \citet{barletal96}], while \citet{zubkelit00} required a
much higher mass-loss rate,
2.3$\times$10$^{-6}$\,M$_{\odot}$\,yr$^{-1}$ (at the larger distance
115\,pc) to explain the ISO H$_2$O data.  We have found that such a
high mass-loss rate produces CO radio lines that are at least a factor
of 30 too strong.  However, the ISO CO $J$=16$\rightarrow$15 and
$J$=17$\rightarrow$16 lines are only about a factor of two too strong. 
Hence, there is some considerable uncertainty in the properties of
this CSE. A fit to the SiO line data using the larger distance and
mass-loss rate is as bad as that for the low distance and mass-loss
rate.

In the case of \object{R~Dor} \citet{olofetal02} could
not fit well the CO radio line profiles.  The model profiles were
sharply double-peaked, while the observed ones were smoothly rounded. 
We merely note here that there was no problem to fit the SiO line profiles
with the nominal values for \object{R~Dor}.

\section{Discussion and conclusions}
\label{s:discussion}

An extensive radiative transfer analysis of circumstellar
SiO `thermal' radio line emission from a large sample of M-type AGB
variable stars have been performed, partly based on a new, large,
observational data base.  It is concluded that, at this stage, the
modelling of the circumstellar SiO radio line emission is considerably more
uncertain than that of the CO radio line emission.  Partly because the
SiO line emission predominantly comes from the inner regions where the
observational constraints are poor, but also partly because the
behaviour of the SiO molecule is more complex, e.g., adsorption onto
grains.  A rather detailed sensitivity analysis has been done, in
order to estimate the reliability of the derived results.

In particular, the size of the SiO envelope is crucial to the
modelling.  Multi-line SiO modelling of eleven sources were used to
establish a relation between the size of the SiO envelope and the
density measure $\dot{M}/v_{\rm e}$.  This is of course rather
uncertain, both in the absolute scale and in the dependence on the
density measure.  Comparison with estimates based on rather simple
condensation and photodissociation theories suggests that the derived
relation is not unreasonable. A very good agreement with the photodissociation
radii is obtained for an unshilded photodissociation rate of 
2.5\,$\times$\,10$^{-10}$\,s$^{-1}$. It was also checked against
interferometeric SiO line brightness size estimates of six sources.

The SiO abundance distribution of the IRV/SRV
sample has a median value of 6$\times$10$^{-6}$, and a minimum of
2$\times$10$^{-6}$ and a maximum of 5$\times$10$^{-5}$.  For these,
low to intermediate mass-loss rate objects, we expect the abundances
to be representative for the region inside the SiO photodissociation
radius.  This applies also to the low and intermediate mass-loss rate Miras. The
high mass-loss rate Miras have a median abundance which is more than a
factor of six lower than that of the IRV/SRV sample.  The
derived SiO abundances are in all cases (within the uncertainties)
below the abundance expected
from stellar atmosphere chemistry ($\approx$\,4$\times$10$^{-5}$,
\citet{duaretal99}), the median for the total sample is down by a factor
of ten.  We regard this as a safe result, and interpret it in terms
of SiO adsorption onto grains, which is efficient already at low mass-loss
rates.

In addition, there is a trend of decrasing SiO abundance with
increasing mass-loss rate, Fig.~\ref{abundancemassloss}.  Here, we
cannot entirely exclude systematic effects of the modelling.  In
particular, the adopted SiO envelope size relation can introduce such
effects.  E.g., smaller envelopes at low mass-loss rates, as indicated
by the photodissociation model, would increase the estimated
abundances for these objects.  This will actually strenghten the
observed trend.  There is no obvious reason for a similar envelope
size decrease for the high mass-loss rate objects (however, see
below), but if present it would lead to a less pronounced trend.  The
discussion in Sect.~\ref{s:dependence} on the sensitivity on the
radiation field distribution suggests that the abundance estimates for
the high mass-loss rate objects are upper limits.  Therefore,
considering also that the effect is rather large, we regard the trend
as at least tentative.  An interpretation in terms of increased
adsorption of SiO onto grains with increasing mass-loss rate is
natural.  In Fig.~\ref{abundancemassloss} a depletion curve based on
the results in Sect.~\ref{s:abudistr} is plotted, and it represents
well the general trend (the adopted parameters are $T_{\rm
bb}$\,=\,2500\,K, $L$\,=\,4000\,L$_{\odot}$,
$v_{\rm e}$\,=\,10\,km\,s$^{-1}$, $\psi$\,=\,0.002, $a_{\rm
gr}$\,=\,0.05\,$\mu$m, $\rho_{\rm gr}$\,=\,2\,g\,cm$^{-3}$, $Q_{\rm
p,F}$\,=\,0.03, $T_{\rm bind,SiO}$\,=\,29500\,K, and $\alpha$\,=\,1). 
We emphasize once again how sensitive the theoretical condensation
results are to the adopted parameters, and the depletion curve can
easily be made to fit better the estimated abundances.

For the high mass-loss rate Miras the SiO abundance distribution appears
bimodal, a low
abundance group (on average 4$\times$10$^{-7}$) and a high abundance
group (on average 5$\times$10$^{-6}$).
At this point we cannot identify any reason for this.  The stars
and their CSEs are rather similar and there is no reason to expect the
modelling to artificially produce very different results for rather
similar objects, but the SiO line modelling of these objects are 
particularly difficult as discussed above.  The high values can be 
explained if there is a
process which decreases the condensation onto, or leads to effective
evaporation from, dust grains for some objects.  The former is
possible if the dust-to-gas mass ratio is low.  If, on the other hand,
the dust-to-gas mass ratio is high, the region contributing to the SiO
line emission may be much smaller than used in our modelling, and
hence the SiO abundance is underestimated.  This may be the case for the low
abundance objects.  Substantial mass-loss rate variations with time may
of course lead to surprising results.  This can possibly be checked by
high angular resolution observations of both CO and SiO radio line emission.
It is also interesting that \citet{woodsetal03b} found, in a
sample of high mass-loss rate C-stars, that the SiO abundance is one
of the few of their abundance estimates that vary significantly from
star to star.  We note, though, that for the C-stars the estimated SiO
abundances are low, about 1$\times$10$^{-7}$, and that
\citet{willcher98} have shown that for C-stars shock
chemistry may significantly alter the SiO abundance.  The same is not
the case for the O-rich chemistry according to 
\citet{duaretal99}, but grains are not included in their analysis.

The SiO and CO radio line profiles differ in shape.  For those stars
with high enough S/N-ratio data on both species, it is clear that the
dominating parts of the SiO profiles are narrower than the CO profiles,
but the former have low-intensity wings which cover the full velocity range
of the CO profile.  The effect is more evident in high-$J$ lines, and
less evident in high mass-loss rate objects.  This is interpreted (as
has been done also by others) as due to the influence of gas
acceleration in the region which produces most of the SiO line
emission.  This points to a weakness in our analysis.  Clearly, this
acceleration region must be treated more carefully in the radiative
modelling, but this is also the region where condensation occurs, a
process which is difficult to describe in detail.  

These results strongly suggest that SiO radio line emission can be
used as a sensitive probe of circumstellar dust formation and
dynamics.  However, considerable progress in this area can only be
expected from a combination of high-quality SiO multi-line
observations, high-quality interferometric observations of a number of
SiO lines for a representative sample of sources, a detailed radiative
transfer analysis, which includes also the dust radiation, and a
detailed SiO chemical model. The rather high $\chi^2$ values of some 
of our best-fit
models suggest that our circumstellar model needs to be improved.

\citet{olofetal02} identified a number of sources
with peculiar CO line profiles, essentially consisting of a narrow
feature centered on a (much) broader feature.  These have been discussed
here from the point of view of their SiO line properties.  Except in
one case, the SiO and CO line profiles are rather similar, and the
derived SiO abundances are in no case peculiar.  The low gas expansion
velocity source L$^2$~Pup has a very narrow SiO line profile as expected, but
also a considerably broader, low-intensity
component.  Finally, W~Hya imposes a problem for the SiO line
modelling.  In principle, a much larger SiO envelope than warranted by
the mass-loss rate derived from the CO data is required to fit well
the (high-quality) SiO data.  This, in combination with other data,
suggest that the CSE of this star is not normal, possibly as an effect
of time-variable mass loss.

\begin{acknowledgements}
Financial support from the Swedish Science Research Council is gratefully
acknowledged by DGD, FLS, ML, and HO. FK's work was supported by APART (Austrian
Programme for Advanced Research and Technology) from the Austrian
Academy of Sciences and by the Austrian Science Fund Project
P14365-PHY. FLS further acknowledges support from the Netherlands Organization
for Scientific Research (NWO) grant 614.041.004.
\end{acknowledgements}



\appendix

\section{Observational results}
    
\begin{table*}
     \centering
     \caption[]{Observational results of circumstellar SiO radio line emission towards 
                a sample of M-type IRVs and SRVs (Part 1)}
      \label{t:srvirvobsresultsI}
    \[
	\begin{tabular}{llrrclrrrcc}
\hline
\noalign{\smallskip}
GCVS4 & IRAS  & Code & S & $I$ & $T_{\rm mb}$ & $v_{\rm hel}${\phantom{00}} & $v_{\rm LSR}${\phantom{00}}
&  $v_{\rm e}${\phantom{00}} & Q & C \\
       &       &      &   & [K\,km\,s$^{-1}$] & [K] & [km\,s$^{-1}$] &
[km\,s$^{-1}$] & [km\,s$^{-1}$] & & \\
\hline
BC And   & 22586+4614 &O21&N&{\phantom{11}}0.2  &       &        &     &     & 5 &\\
RS And   & 23528+4821 &O21&D&{\phantom{11}}0.71 & 0.076 & $-$3.1 & 4.2 & 6.1 & 3 &\\
UX And   & 02302+4525 &O21&D&{\phantom{11}}0.94 & 0.044 & $-$19.4 & $-$20.1 & 16.6 & 3 &\\
$\theta$ Aps & 14003--7633 &S21&D&{\phantom{11}}0.62 & 0.078 & 10.8 & 3.1 & 6.1 & 2 &\\
          &                &S32&D&{\phantom{11}}1.4  & 0.18  & 10.9 & 3.2 & 5.6 & 2 &\\
          &                &S54&D&{\phantom{11}}2.7  & 0.30  & 10.7 & 3.0 & 7.6 & 2 &\\          
TZ Aql   & 20276--0455&S21&D&{\phantom{11}}0.15 & 0.025 & 49.3 & 62.3 & 5.0 & 3 &\\
          &           &S32&D&{\phantom{11}}0.14 & 0.026 & 49.4 & 62.4 & 4.5 & 4 &\\
V584 Aql & 20079--0146&S21&N&{\phantom{11}}0.2  &       &        &     &     & 5 &\\
          &           &S32&N&{\phantom{11}}0.1  &       &        &     &     & 5 & \\
AB Aqr   & 22359--1417&S21&N&{\phantom{11}}0.1  &       &        &     &     & 5 &\\
          &           &S32&N&{\phantom{11}}0.1  &       &        &     &     & 5 & \\	  
EP Aqr   & 21439--0226&S21&D&{\phantom{11}}3.2 & 0.28 & $-$41.3 & $-$31.9 & 8.2 & 2 &\\
          &           &S32&D&{\phantom{11}}4.4 & 0.40 & $-$41.3 & $-$31.9 & 7.9 & 2 &\\
SV Aqr   & 23201--1105&S21&D&{\phantom{11}}0.54 & 0.036 & 7.3 & 8.5 & 9.5 & 3 &\\
          &           &S32&D&{\phantom{11}}0.66 & 0.070 & 6.6 & 7.8 & 7.2 & 2 &\\
T Ari    & 02455+1718 &O21&D&{\phantom{11}}0.084 & 0.043 & 6.5 & $-$1.2 & 1.5 & 4 &\\
          &           &S21&D&{\phantom{11}}0.081 & 0.021 & 6.4 & $-$1.4 & 2.9 & 4 &\\
          &           &S32&D&{\phantom{11}}0.091 & 0.017 & 5.5 & $-$2.2 & 4.0 & 4 &\\
RV Boo   & 14371+3245 &O21&D&{\phantom{11}}0.79 & 0.069 & $-$5.0 & 9.5 & 6.5 &&b\\
	 &            &O21&D&{\phantom{11}}0.19 & 0.051 & $-$6.0 & 8.5 & 2.0 &&n\\
	 &            &O21&D&{\phantom{11}}0.98 & 0.12  & $-$6.9  & 7.6  & 8.8 & 2 &b+n\\
	 &            &I32&D&{\phantom{11}}2.5 & 0.25 & $-$5.9 & 8.6 & 8.0 &&b+n\\	 
RX Boo   & 14219+2555 &O21&D&{\phantom{11}}9.5 & 0.84 & $-$12.1 & 0.2 & 9.0 & 1 &\\
          &           &S21&D&{\phantom{11}}5.1 & 0.44 & $-$11.6 & 0.7 & 8.7 & 1 &\\
          &           &S32&D&{\phantom{11}}6.4 & 0.61 & $-$11.7 & 0.6 & 8.2 & 1 &\\
RV Cam   & 04265+5718 &O21&D&{\phantom{11}}0.32 & 0.036 & $-$15.9 & $-$16.0 & 6.8 & 4 &\\
BI Car  & 10416--6313 &S21&N&{\phantom{11}}0.2  &       &        &     &     & 5 & \\
          &           &S32&N&{\phantom{11}}0.3  &       &        &     &     & 5 & \\
SS Cep   & 03415+8010 &O21&N&{\phantom{11}}0.3  &       &        &     &     & 5 & \\
V744 Cen & 13368--4941 &S21&N&{\phantom{11}}0.1  &       &        &     &     & 5 & \\
          &            &S32&N&{\phantom{11}}0.4  &       &        &     &     & 5 & \\
V806 Cen & 13465--3412 &S21&N&{\phantom{11}}0.3  &       &        &     &     & 5 & \\
          &            &S32&N&{\phantom{11}}0.2  &       &        &     &     & 5 & \\	  
UY Cet   & 00245--0652&O21&D&{\phantom{11}}0.41 & 0.071 & 6.5 & 3.6 & 4.6 & 4 &\\
          &           &S21&D&{\phantom{11}}0.38 & 0.053 & 7.8 & 4.9 & 5.2 & 4 &\\
          &           &S32&D&{\phantom{11}}1.02 & 0.11  & 8.9 & 6.0 & 7.1 & 3 &\\
CW Cnc   & 09057+1325 &O21&D&{\phantom{11}}0.62 & 0.065 & 26.6 & 17.2 & 7.9 & 3 &\\
R Crt    & 10580--1803&S21&D&{\phantom{11}}5.0 & 0.33 & 19.6 & 11.9 & 10.9 & 1 &\\
          &           &S54&D&{\phantom{11}}9.9 & 0.66 & 19.8 & 12.1 & 10.7 & 2 &\\
          &           &S65&D&{\phantom{1}}13.6 & 0.86 & 19.0 & 11.3 & 12.6 & 3 &\\
V CVn    & 13172+4547 &O21&N&{\phantom{11}}0.3  &       &        &     &     & 5 & \\
W Cyg    & 21341+4508 &O21&N&{\phantom{11}}0.1  &       &        &     &     & 5 & \\
U Del    & 20431+1754 &O21&N&{\phantom{11}}0.5  &       &        &     &     & 5 & \\
R Dor    & 04361--6210&S21&D&{\phantom{11}}9.6  & 1.1 & 23.9 & 7.5 & 6.1 & 1 &\\
          &           &S32&D&{\phantom{1}}24.0  & 2.8 & 24.0 & 7.6 & 6.0 & 1 &\\
          &           &S54&D&{\phantom{1}}31.1  & 3.6 & 24.4 & 8.0 & 6.0 & 1 &\\
          &           &S65&D&{\phantom{1}}34.7  & 5.1 & 24.4 & 8.0 & 5.8 & 1 &\\
AH Dra   & 16473+5753 &I32&D&{\phantom{11}}1.2 & 0.17 & 57.6 & 74.9 & 8.5 &&\\	  
CS Dra   & 11125+7524 &O21&D&{\phantom{11}}0.41 & 0.022 & $-$60.1 & $-$51.3 & 12.5 & 4 &\\
S Dra    & 16418+5459 &O21&D&{\phantom{11}}1.4 & 0.10 & $-$2.2 & 15.4 & 9.9 & 4 &\\
          &           &I32&D&{\phantom{11}}3.6 & 0.30 & $-$1.6 & 15.9 & 8.6 & 3 &\\
SZ Dra   & 19089+6601 &O21&D&{\phantom{11}}0.17 & 0.024 & $-$44.2 & $-$28.3 & 4.8 & 4 &\\
TY Dra   & 17361+5746 &O21&D&{\phantom{11}}0.65 & 0.038 & $-$33.9 & $-$18.9 & 11.5 & 4 &\\
          &           &I32&D&{\phantom{11}}3.2  & 0.22  & $-$34.0 & $-$16.4 & 11.1 & 3 &\\
UU Dra   & 20248+7505 &O21&N&{\phantom{11}}0.3  &       &        &     &     & 5 & \\
g Her    & 16269+4159 &O21&N&{\phantom{11}}0.2  &       &        &     &     & 5 & \\
X Her    & 16011+4722 &O21&D&{\phantom{11}}2.2  & 0.18 & $-$90.5 & $-$73.0 & 8.0 &&b\\
	 &            &O21&D&{\phantom{11}}0.53 & 0.16 & $-$89.5 & $-$72.0 & 2.2 &&n\\
	 &            &O21&D&{\phantom{11}}2.7  & 0.34 & $-$90.7 & $-$73.2 & 7.4 & 1 &b+n\\
FZ Hya   & 08189+0507 &O21&N&{\phantom{11}}0.7  &       &        &     &     & 5 & \\
\hline
\noalign{\smallskip}
\end{tabular}
         \]
\end{table*}

\begin{table*}
     \centering
     \caption[]{Observational results of circumstellar SiO radio line emission towards 
a sample of M-type IRVs and SRVs (Part 2)}
      \label{t:srvirvobsresultsII}
    \[
	\begin{tabular}{llrrclrrrcc}
\hline
\noalign{\smallskip}
GCVS4 & IRAS  & Code & S & $I$ & $T_{\rm mb}$ & $v_{\rm hel}${\phantom{00}} & $v_{\rm LSR}${\phantom{00}}
&  $v_{\rm e}${\phantom{00}} & Q & C \\
       &       &      &   & [K\,km\,s$^{-1}$] & [K] & [km\,s$^{-1}$] &
[km\,s$^{-1}$] & [km\,s$^{-1}$] & & \\
\hline
W Hya    & 13462--2807&S21&D&{\phantom{11}}6.0 & 0.63 & 38.5 & 40.4 & 7.1 & 1 &\\
          &           &S32&D&{\phantom{1}}10.9 & 1.2  & 39.2 & 41.1 & 6.4 & 1 &\\
          &           &S54&D&{\phantom{1}}13.2 & 0.98 & 39.2 & 41.1 & 7.1 & 1 &\\
          &           &S65&D&{\phantom{1}}13.2 & 0.86 & 39.4 & 41.3 & 6.7 & 1 &\\
RW Lep   & 05365--1404&S21&N&{\phantom{11}}0.3  &       &        &     &     & 5 & \\
         &            &S32&N&{\phantom{11}}0.1  &       &        &     &     & 5 & \\
U Men    & 04140--8158&S21&D&{\phantom{11}}0.30 & 0.037 & 28.9 & 17.0 & 6.3 & 3 &\\
          &           &S32&D&{\phantom{11}}0.44 & 0.049 & 28.7 & 16.8 & 6.5 & 3 &\\
T Mic    & 20248--2825&S21&D&{\phantom{11}}0.67 & 0.088 & 17.7 & 25.3 & 5.6 & 2 &\\
          &           &S32&D&{\phantom{11}}1.2  & 0.16  & 17.8 & 25.4 & 5.6 & 2 &\\
EX Ori   & 05220--0611&O21&N&{\phantom{11}}0.7  &       &        &     &     & 5 & \\
V352 Ori & 05592--0221&O21&N&{\phantom{11}}0.5  &       &        &     &     & 5 & \\
S Pav    & 19510--5919&S21&D&{\phantom{11}}0.32 & 0.030 & $-$19.2 & $-$20.0 & 8.3 & 3 &\\
          &           &S32&D&{\phantom{11}}0.40 & 0.055 & $-$19.1 & $-$19.9 & 5.6 & 4 &\\
NU Pav   & 19575--5930&S21&N&{\phantom{11}}0.1  &       &        &     &     & 5 & \\
          &           &S32&N&{\phantom{11}}0.3  &       &        &     &     & 5 & \\	  
SV Peg   & 22035+3506 &O21&D&{\phantom{11}}2.1 & 0.18 & $-$8.3 & 2.1 & 9.5 & 2 &\\
TW Peg   & 22017+2806 &O21&D&{\phantom{11}}0.54 & 0.035 & $-$23.0 & $-$10.5 & 9.3 & 4 &\\
SV Psc   & 01438+1850 &O21&D&{\phantom{11}}0.45 & 0.05  & 9.5  & 5.7  & 6.9  & 3 & b+n\\
         &            &S21&D&{\phantom{11}}0.31 & 0.03  & 10.5  & 6.7  & 11.0  & 3 & b+n\\
         &            &S54&D&{\phantom{11}}0.63 & 0.06  & 10.4  & 6.6  & 8.4  & 3 & b \\
         &            &S54&D&{\phantom{11}}0.20 & 0.10  & 10.4  & 6.6  & 1.6  & 3 & n \\
         &            &S54&D&{\phantom{11}}0.83 & 0.10  & 10.2  & 6.4  & 6.3  & 3 & b+n \\                           
V PsA    & 22525--2952&S21&N&{\phantom{11}}0.1  &       &        &     &     & 5 & \\
          &           &S32&N&{\phantom{11}}0.1  &       &        &     &     & 5 & \\
L$^2$ Pup& 07120--4433&S21&D&{\phantom{11}}0.83 & 0.14 & 52.4 & 33.6 & 4.6 & 2 &\\
          &           &S32&D&{\phantom{11}}1.7  & 0.30 & 52.0 & 33.3 & 3.5 & 2 &\\
          &           &S54&D&{\phantom{11}}2.5  & 0.54 & 52.3 & 33.6 & 3.6 & 1 &\\          
Y Scl    & 23063--3024&S21&D&{\phantom{11}}0.22 & 0.024 & 30.9 & 29.5 & 6.4 & 4 &\\
          &           &S32&D&{\phantom{11}}0.28 & 0.036 & 31.8 & 30.4 & 5.8 & 3 &\\
CZ Ser   & 18347--0241&O21&N&{\phantom{11}}0.4  &       &        &     &     & 5 & \\	  
$\tau^4$Ser&15341+1515&O21&N&{\phantom{11}}0.2  &       &        &     &     & 5 & \\
          &           &S21&N&{\phantom{11}}0.2  &       &        &     &     & 5 & \\
          &           &S32&N&{\phantom{11}}0.1  &       &        &     &     & 5 & \\
SU Sgr   & 19007--2247&S21&N&{\phantom{11}}0.1  &       &        &     &     & 5 & \\
          &           &S32&N&{\phantom{11}}0.2  &       &        &     &     & 5 & \\
V1943 Sgr& 20038--2722&S21&D&{\phantom{11}}0.96 & 0.13 & $-$23.1 & $-$14.4 & 5.4 & 2 &\\
          &           &S32&D&{\phantom{11}}1.7  & 0.26 & $-$23.0 & $-$14.3 & 4.8 & 2 &\\
V Tel    & 19143--5032&S21&D&{\phantom{11}}0.59 & 0.063 & $-$34.6 & $-$31.8 & 7.1 &3 &\\
          &           &S32&D&{\phantom{11}}0.40 & 0.060 & $-$35.0 & $-$32.2 & 5.4 & 3 &\\
Y Tel    & 20165--5051&S21&D&{\phantom{11}}0.21 & 0.034 & $-$46.7 & $-$45.3 & 3.8 & 4 &\\
          &           &S32&D&{\phantom{11}}0.44 & 0.066 & $-$45.8 & $-$44.4 & 4.6 & 3 & \\
AZ UMa   & 11445+4344 &O21&N&{\phantom{11}}0.2  &       &        &     &     & 5 & \\
Y UMa    & 12380+5607 &O21&D&{\phantom{11}}1.2  & 0.14 & 8.8 & 18.6 & 6.0 & 2 &\\
SU Vel   & 09480--4147&S21&D&{\phantom{11}}0.35 & 0.035 & 20.8 & 7.0 & 7.4 & 3 &\\
          &           &S32&D&{\phantom{11}}0.38 & 0.041 & 20.8 & 7.0 & 7.4 & 3 &\\
BK Vir   & 12277+0441 &O21&D&{\phantom{11}}0.63 & 0.10 & 15.1 & 17.8 & 4.7 & 3 &\\
RT Vir   & 13001+0527 &O21&D&{\phantom{11}}3.8  & 0.39 & 13.5 & 18.7 & 7.4 & 2 &\\
          &           &S21&D&{\phantom{11}}3.2  & 0.27 & 13.4 & 18.6 & 9.2 & 2 &\\
RW Vir   & 12046--0629&O21&N&{\phantom{11}}0.4  &       &        &     &     & 5 & \\
SW Vir   & 13114--0232&O21&D&{\phantom{11}}3.5  & 0.33 & $-$14.5 & $-$9.8 & 7.9 & 2 &\\
\hline
\noalign{\smallskip}
\end{tabular}
         \]
\end{table*}

\begin{table*}
    \centering
    \caption[]{Observational results of circumstellar CO and SiO line emission 
              towards our Mira sample}
     \label{t:mirasobsresults}
   \[
	\begin{tabular}{llcrcccc}
\hline
\noalign{\smallskip}
GCVS4 & IRAS  & Molecule &Code & $I$ & $T_{\rm mb}$ & $v_{\rm LSR}$ &  $v_{\rm e}$ \\
      &       &    &     & [K\,km\,s$^{-1}$] & [K]  & [km\,s$^{-1}$] & [km\,s$^{-1}$] \\
\hline
TX Cam   & 04566+5606 &CO &O10& {\phantom{1}}21.5 & 0.88 & 11.5 & 20.0 \\
         &            &   &I10& {\phantom{1}}67.0 & 1.44 & 11.5 & 21.0 \\
         &            &   &I21&             241.3 & 3.93 & 11.4 & 20.7 \\
         &            &   &J21& {\phantom{1}}69.7 & 2.61 & 11.4 & 20.2 \\
         &            &   &J32&             167.2 & 6.14 & 11.3 & 20.3 \\
         &            &SiO&O21& {\phantom{1}}13.2 & 0.52 & 10.7 & 18.6 \\

R Cas    & 23558+5106 &CO &O10& {\phantom{11}}8.4 & 0.46 & 24.9 & 12.2 \\
         &            &   &J21& {\phantom{1}}32.1 & 1.78 & 24.8 & 11.1 \\
         &            &   &J32& 	    100.2 & 5.48 & 25.2 & 11.0 \\
         &            &   &J43& {\phantom{1}}89.6 & 5.44 & 24.8 & 11.9 \\
         &            &SiO&O21& {\phantom{11}}8.6 & 0.67 & 26.3 & 9.4 \\

R Hya    & 13269--2301&CO &S10& {\phantom{11}}0.6 & 0.09: & -9.7: & 5.3: \\
	     &            &   &J32& {\phantom{1}}43.0 & 3.65 & $-$10.5 & 8.6 \\
	     & 	          &SiO&S21& {\phantom{11}}2.8 & 0.44 & $-$11.5 & 5.2 \\
	     & 	          &   &S54& {\phantom{11}}7.7 & 1.30 & $-$11.2 & 5.9 \\	     

R Leo    & 09448+1139 &CO &O10& {\phantom{11}}2.4 & 0.25 & 0.2 & 7.4\\
         &            &   &J21& {\phantom{1}}15.0 & 1.30 & $-$0.4 & 8.7 \\
         &            &   &J32& {\phantom{1}}41.6 & 3.88 & $-$0.3 & 7.8 \\
         &            &SiO&O21& {\phantom{11}}5.1 & 0.51 & 0.5 & 6.8 \\
         &            &   &S21& {\phantom{11}}4.9 & 0.65 & 0.2 & 6.9 \\
	     & 	          &   &S54& {\phantom{1}}10.8 & 1.58 & 0.4 & 6.3 \\         

GX Mon   & 06500+0829 &CO &O10& {\phantom{1}}49.6 & 1.55 & $-$9.3 & 24.4 \\
	     &	          &   &J21& {\phantom{1}}67.2 & 2.27 & $-$9.5 & 18.2 \\
         &            &   &J32& {\phantom{1}}87.6 & 2.96 & $-$9.4 & 17.9 \\
         &            &   &J43& {\phantom{1}}56.2 & 1.82 & $-$8.7 & 17.5 \\
         &            &SiO&O21& {\phantom{11}}9.4 & 0.36 & $-$9.6 & 19.5 \\
         &            &   &S21& {\phantom{11}}5.5 & 0.29 & $-$9.6 & 18.9 \\
         &            &   &S54& {\phantom{11}}8.9 & 0.42 & $-$9.1 & 18.7 \\                  

WX Psc   & 01037+1219 &CO &O10& {\phantom{1}}52.0 & 1.61 & 9.6 & 18.6 \\
	     &	          &   &J21& {\phantom{1}}31.0 & 1.31 & 10.1 & 18.3 \\
         &            &   &J32& {\phantom{1}}45.5 & 1.63 & 9.5 & 20.8 \\
         &            &   &J43& {\phantom{1}}49.9 & 1.82 & 9.5 & 20.6 \\
	     &            &SiO&O21& {\phantom{1}}10.0 & 0.40 & 10.1 & 18.9 \\
         &            &   &S21& {\phantom{11}}5.6 & 0.22 & 9.3 & 19.2 \\ 
         &            &   &S54& {\phantom{11}}6.8 & 0.29 & 9.3 & 17.7 \\          	     
IK Tau   & 03507+1115 &CO &O10& {\phantom{1}}58.6 & 1.70 & 34.4 & 17.2 \\
	     &	          &   &J21& 	    103.8 & 3.32 & 34.5 & 17.5 \\
         &            &   &J32& 	    143.8 & 4.81 & 34.2 & 17.6 \\
         &            &   &J43& 	    127.0 & 4.52 & 33.2 & 18.0 \\
	     & 	          &SiO&O21& {\phantom{1}}16.1 & 0.66 & 33.9 & 18.1 \\
         &            &   &S54& {\phantom{1}}18.9 & 0.73 & 34.7 & 17.5 \\
         &            &   &S65& {\phantom{1}}13.7 & 0.61 & 34.7 & 15.9 \\

IRC+10365& 18349+1023 &CO &O10& {\phantom{1}}21.5 & 0.79 & $-$31.3 & 16.9 \\
         &            &   &J21& {\phantom{1}}42.0 & 1.48 & $-$30.7 & 15.3 \\
         &            &SiO&O21& {\phantom{11}}6.3 & 0.30 & $-$32.6 & 15.6 \\

IRC--10529& 20077--0625&CO &O10& {\phantom{1}}15.7 & 0.65 & $-$18.1 & 15.7 \\
         &            &   &J21& {\phantom{1}}24.4 & 1.18 & $-$17.5 & 17.7 \\
         &            &   &J32& {\phantom{1}}48.0 & 2.22 & $-$17.1 & 14.2 \\
         &            &   &J43& {\phantom{1}}57.4 & 2.87 & $-$17.2 & 16.0 \\
	 &	      &SiO&O21& {\phantom{11}}2.9 & 0.16 & $-$17.7 & 14.6 \\

IRC--30398& 18560--2954&CO &J21& {\phantom{1}}44.6& 1.76 & $-$6.4 & 19.4 \\
	 &	       &SiO&S21& {\phantom{11}}0.8& 0.06: & $-$10.1: & 15.3:\\

IRC+40004& 00042+4248 &CO &O10& {\phantom{1}}24.8 & 0.82 & $-$20.5 & 19.2 \\
         &            &   &J21& {\phantom{1}}41.5 & 1.41 & $-$20.5 & 18.9 \\
  	 &	      &SiO&O21& {\phantom{11}}1.3 & 0.05 & $-$21.3 & 20.3 \\

IRC+50137& 05073+5248 &CO &J21& {\phantom{1}}36.8 & 1.40 & 3.5 & 18.7 \\
         &            &   &J32& {\phantom{1}}35.7 & 1.32 & 3.3 & 18.1 \\
         &	      &SiO&O21& {\phantom{11}}2.0 & 0.10 & 1.4 & 14.4 \\
\hline
\noalign{\smallskip}
\end{tabular}
        \]
\end{table*}

\section{Spectra}

\begin{figure*}
     \centering \includegraphics[width=18cm]{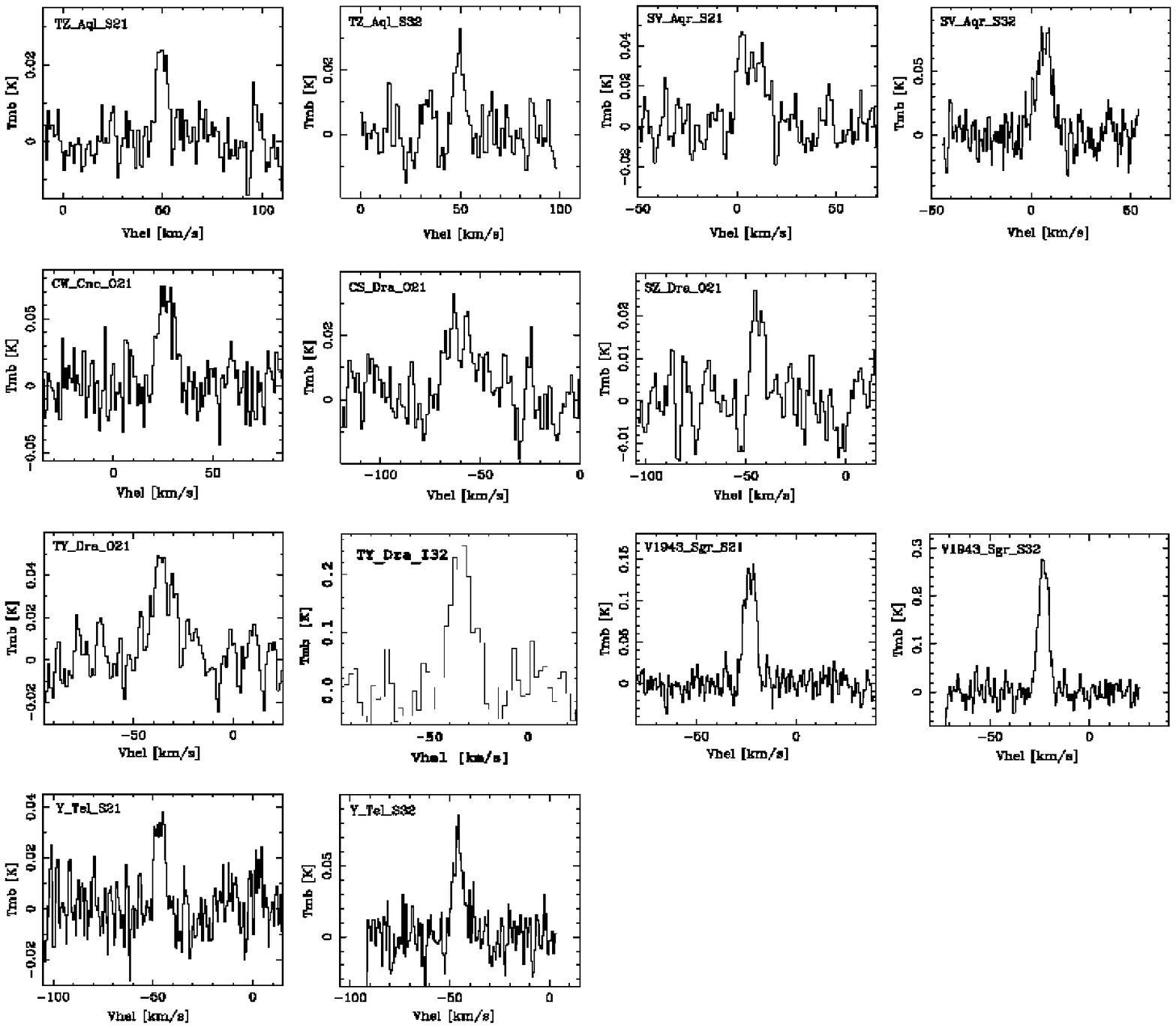}
     \caption{SiO spectra of M-type IRVs. Note the heliocentric velocity scale}
     \label{f:siolb}
\end{figure*}

\begin{figure*}
     \centering \includegraphics[width=18cm]{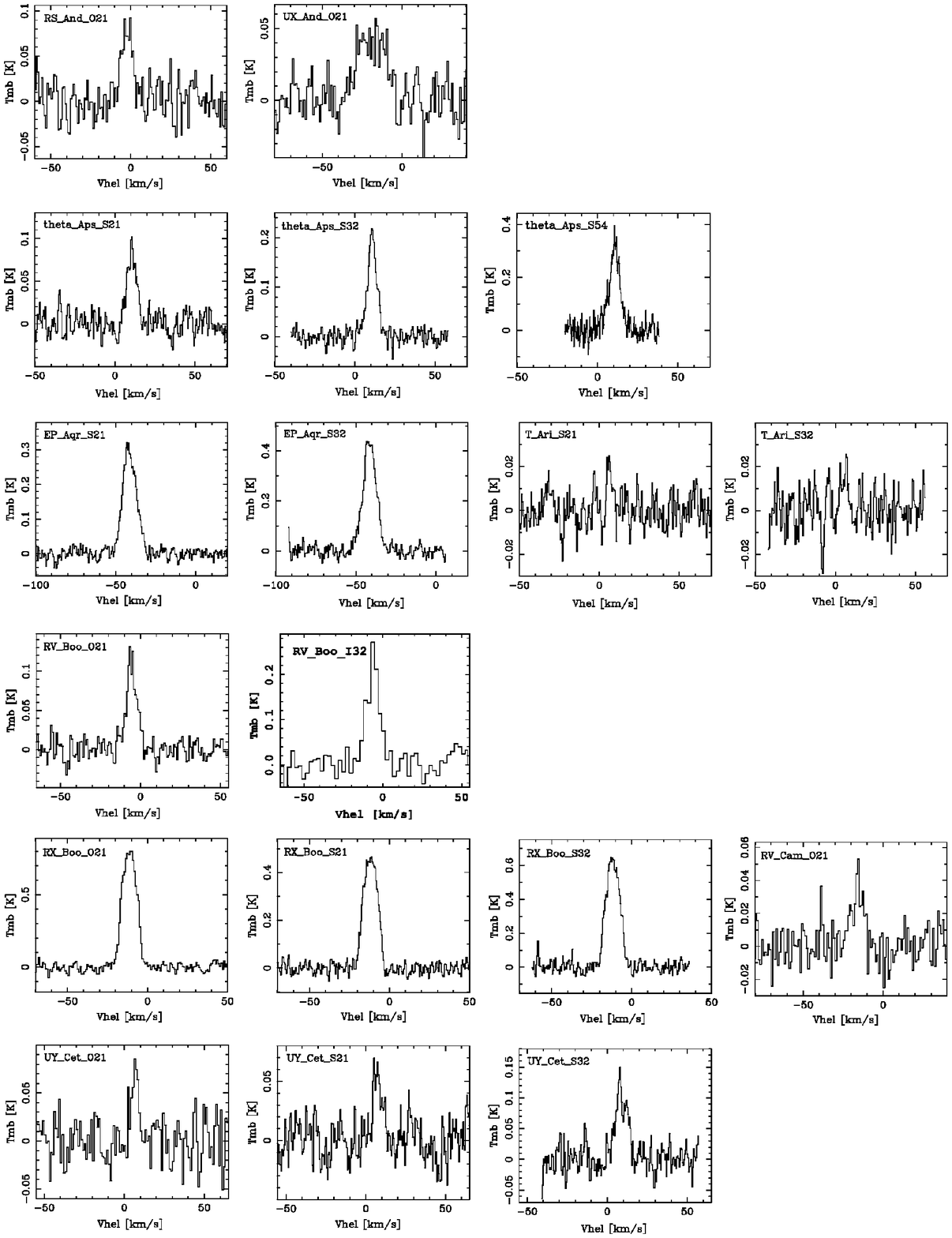}
     \caption{SiO spectra of M-type SRVs (Part 1). Note the heliocentric velocity scale}
     \label{f:siosr1}
\end{figure*}

\begin{figure*}
     \centering \includegraphics[width=18cm]{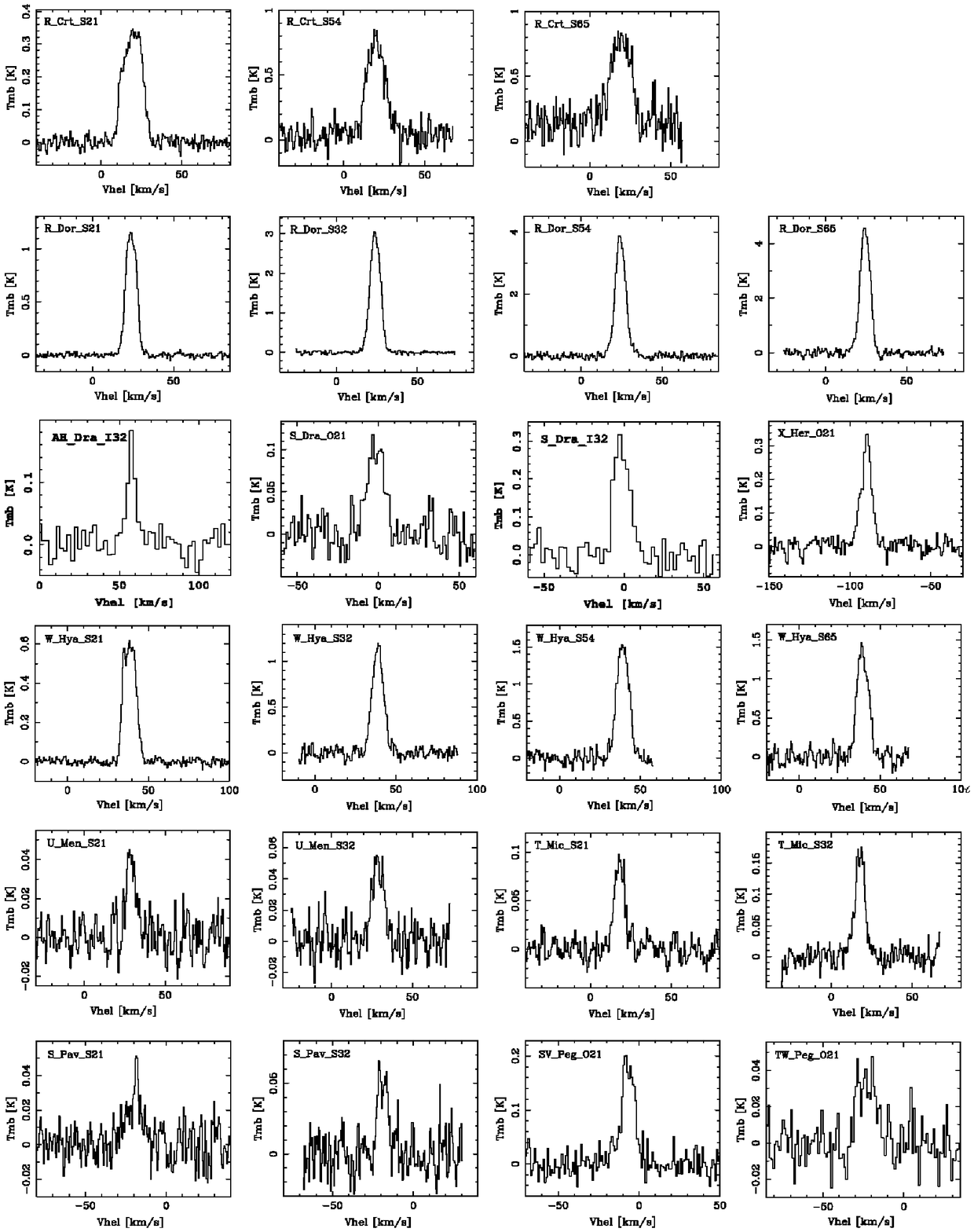}
     \caption{SiO spectra of M-type SRVs (Part 2). Note the heliocentric velocity scale}
     \label{f:siosr2}
\end{figure*}

\begin{figure*}
     \centering \includegraphics[width=18cm]{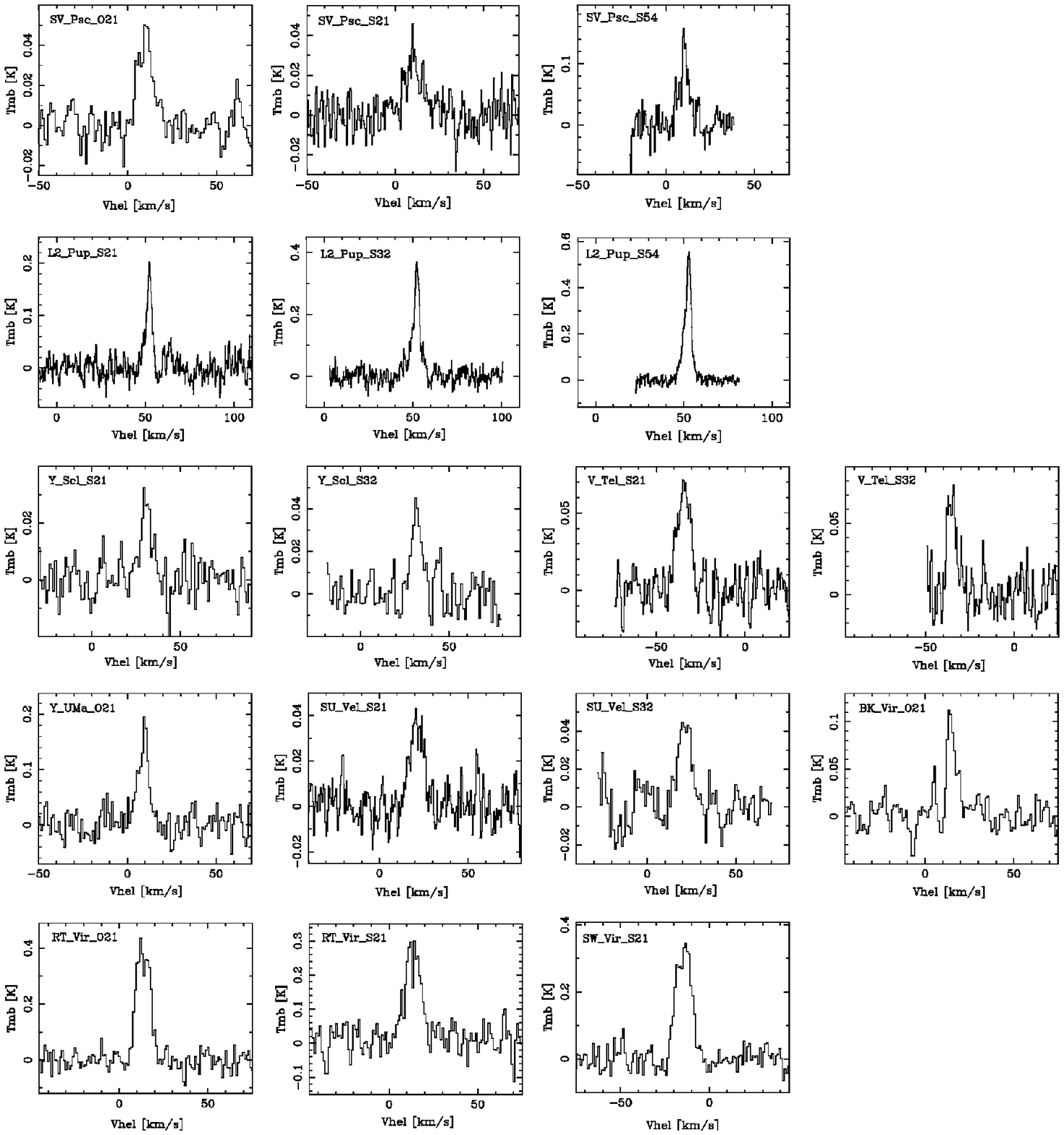}
     \caption{SiO spectra of M-type SRVs (Part 3). Note the heliocentric velocity scale}
     \label{f:siosr3}
\end{figure*}

\begin{figure*}
     \centering \includegraphics[width=18cm]{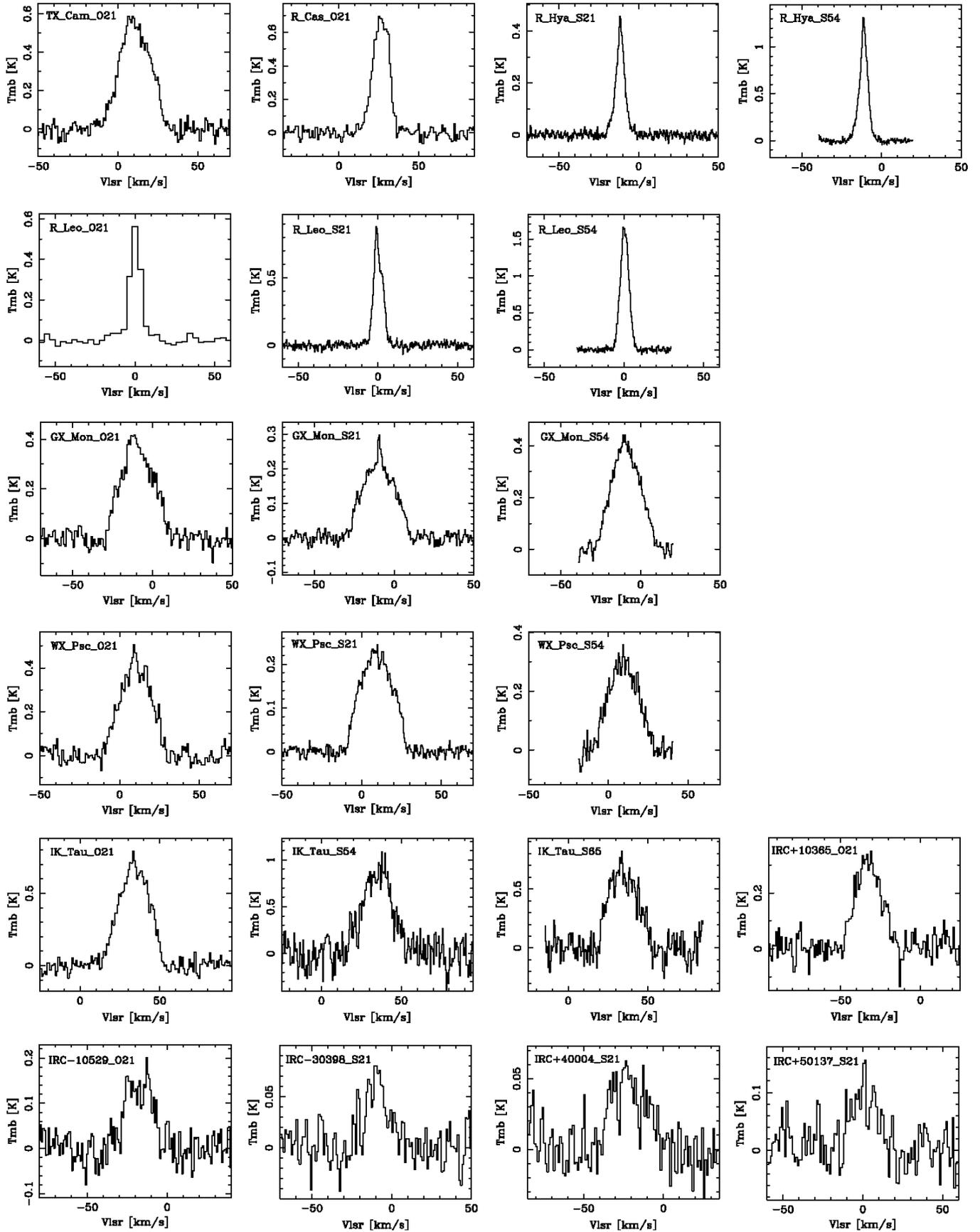}
     \caption{SiO spectra of M-type Miras. Note the LSR
      velocity scale}
     \label{f:siomira}
\end{figure*}

\end{document}